\documentclass[prb,twocolumn,superscriptaddress,showpacs,amsmath,amssymb]{revtex4}
\usepackage{amssymb}
\usepackage{graphicx}
\usepackage{amsmath}
\usepackage{bm}
\usepackage{verbatim}

\begin{document}
\title{Phase transitions in the domain structure of ferromagnetic superconductors.}
%%% author(s) ( + e-mail)

\author{ I.M.\ Khaymovich}
\affiliation{ O.V. Lounasmaa Laboratory, Aalto University, P.O. Box 15100, 00076 Aalto, Finland}
\affiliation{Institute for Physics of Microstructures, Russian
Academy of Sciences, 603950 Nizhny Novgorod, GSP-105, Russia }
\author{A.S.\ Mel'nikov}
\affiliation{Institute for Physics of Microstructures, Russian
Academy of Sciences, 603950 Nizhny Novgorod, GSP-105, Russia }
\affiliation{Lobachevsky State University of Nizhni Novgorod, 23 Prospekt Gagarina, 603950, Nizhni Novgorod, Russia}

\author{A.I.\ Buzdin}
\affiliation{Universit\'e Bordeaux and Institut Universitaire de France, LOMA, UMR 5798, F-33400 Talence, France}

\date{\today}
%%% dates of submition & resubmition (if submitted once, second argument is *)
%%% abstract

\begin{abstract}
Starting from the London - type model we study the domain structures in ferromagnetic superconductors taking account of the nucleation of vortices and antivortices coupled to the magnetic texture.
We predict that the coupling between domains and vortices results in the formation of two
 energetically favorable domain configurations: (i) a Meissner - type vortex free configuration with strong domain shrinking and (ii) a more rare domain configuration with a dense vortex -- antivortex lattice. The switching between these configurations is shown to result in the first order phase transitions which could be observable in superconducting uranium based compounds.
\end{abstract}
%%% PACS numbers
\maketitle
%\renewcommand{\@dotsep}{4.5}
%\tableofcontents

%%%%%%%%%%%%%%%%%%%%%%%%%%%%%%%%%%%%%%%%%%%%%%%%%%%%%%%%%%%%%%%%%%%%%%%%%%%%%%%%%%%%
\section{Introduction}\label{sec1_intro}
The first two ferromagnetic superconductors (FS) $UGe_2$ and  $URhGe$ were discovered at the beginning of this millennium [\onlinecite{SCferro-1,SCferro-2}] and later the third FS  $UCoGe$ joined this list [\onlinecite{UCoGe-1}]. Their Curie temperature $\theta$ is substantially higher than the superconducting (SC) critical temperature $T_c$, which evidences the triplet character of the superconductivity. Indeed, due to the high exchange field acting on the electron spins in the ferromagnet the singlet superconductivity is incompatible with the ferromagnetism (see [\onlinecite{BuzdBulaevKulic-review-1, KulicBuzd-review-2}] for review).
In the FS, as $\theta> T_c$ , the superconductivity appears in the ferromagnetic state where usually a domain structure exists. The presence of these magnetic domains has been revealed, for example, in the unusual temperature dependence of the upper critical field near $T_c$ [\onlinecite{Aoki-Floquet-1}] and recently they were directly observed  by scanning SQUID microscopy in $UCoGe$ [\onlinecite{UCoGe-2_Paulsen-2012}].  The interaction between magnetic induction and superconductivity may strongly influence the properties of the domain structure and can even cause an intrinsic domain structure generation which was first addressed by Krey [\onlinecite{Krey}].
In the limit $T_c <\theta$ the domain wall energy is too large and this prevents the formation of the intrinsic domain structure [\onlinecite{BulaevBuzdCrotov-1983}].  However, due to the demagnetization effect the domains structure is inherent to the majority of the ferromagnetic films. An interesting question how the superconductivity should modify the equilibrium size of the domain structure was considered in [\onlinecite{Sonin, Buzd_Faure, Sonin_comment_BuzdinFaure_reply, Buzd_Dao}] for the case when the superconductor is in the Meissner state.
In these papers it has been shown that with decreasing temperature the domain size in FS firstly shrinks essentially, while for further decrease in the penetration depth $\lambda$ the monodomain state becomes more energetically favorable.
Besides the vortices penetrating the sample may change the equilibrium domain size and in the limit $\lambda\to 0$
one can get that the domain size decreases with the increase in the magnetization amplitude (see, e.g., Ref.~\onlinecite{Sonin}).
In the present work we analyze the domain structure in the vortex state and determine the conditions of the transition between Meissner and vortex phases for arbirtary penetration depth values.

In recent years a lot of attention has been paid to the magnetism and superconductivity interplay in superconductor-ferromagnet heterostructures (see, e.g., [\onlinecite{Pokrovsky,Buzd-review,AladMoshch-review}] and references therein). When a thin oxide layer separates superconductor and ferromagnet the only mechanism of their interaction is electromagnetic similar to the triplet FS.  Nevertheless the physical consequences of this interaction in heterostructures are very different from FS where the domain size shrinking can achieve several orders of magnitude. In particular,
for superconductor-ferromagnet bilayers the electromagnetic interaction may lead to a maximum $~15\%$ contraction of the ferromagnetic domains in equilibrium state as it has been first predicted by Genkin, Tokman, and Skuzovatkin in Ref.~\onlinecite{Genkin_Tokman} (see also later analysis in
Refs.\
[\onlinecite{BulaevChudn_SF-bilayer, Stankiewicz}]). If one takes into account vortex pinning potentials the expansion of the domains at low temperatures takes place and the application of ac magnetic field, routinely used for equilibration of
domains, may lead to their significant contraction. (see, e.g., [\onlinecite{Meln_Buzdin,dynamical_pinning_SF-2}]).

Of course, the electromagnetic mechanism of domain structure
modification should be very sensitive to a possible nucleation of
vortices and antivortices at the domain boundaries. Recently
isolated vortices and antivortices generated by the magnetic
domains were nicely observed by the scanning force microscopy in
the superconductor-ferromagnet bilayers [\onlinecite{Iavarone}].
The coupling between magnetic domains and vortices may result in
the interesting dynamical effect: an oscillating magnetic field
combining with vortex pinning  leads to an important contraction
of the domains mentioned above
[\onlinecite{dynamical_pinning_SF-1,Meln_Buzdin,dynamical_pinning_SF-2}].
The supercurrents induced by the domain structure are responsible
for a certain pinning potential profile acting on vortices and at
the same time vortex distribution itself plays the role of pinning
for the domain boundaries. It is naturally to expect that such
mutual pinning phenomena should cause a variety of the hysteretic
phenomena in the system, which are, in fact, analogous to the
so-called ``field-cooled'' and ``zero field cooled'' phenomena in
usual superconductors. In other words we can expect that for a
given temperature there may exist two domain structure
configurations which are stable with respect to rather small
perturbations of the domain size and/or the vortex concentration.
One of these configurations corresponds to the Meissner state of
the superconducting subsystem, while the other one contains a
rather dense vortex lattice with the characteristic intervortex
distance small compared to the domain size. The changing of the
temperature modifies the minimal energy value for each of these
configurations and, as a consequence, we can get the switching
between configurations. This switching is accompanied by the
abrupt changing of domain size of the system with eventual
formation (or annihilation) of dense vortex-antivortex lattices in
the adjusting domains. Such switching effects should result in the
first order phase transitions
 and are the most prominent in the ferromagnetic superconductors, where the domain size shrinking in the Meissner state can be of order of magnitude while penetration of a rather large number of vortices into the sample naturally restores the domain structure inherent to the nonsuperconducting state.
The goal of the present paper is to suggest a theoretical description of these first order phase transitions which we believe to be observable in uranium based compounds.

In the present work we neglect the effect of intrinsic pinning of
both  magnetic domain walls and vortices at the inhomogeneities.
In real FS compounds these pinning effects may be, of course,
rather strong. For example, the recent magnetization measurements
in $UCoGe$ [\onlinecite{UCoGe-2_Paulsen-2012}] show an important
vortex  pinning effect, overcoming the  magnetic domain pinning.
The interplay between these two types of pinning makes the physics
of the domain structure in FS very reach. We believe, however,
that the method of the applying a weak oscillating magnetic field,
similar to the one used in
[\onlinecite{dynamical_pinning_SF-1,Meln_Buzdin,dynamical_pinning_SF-2}],
should permit to attend the vortex states close to the equilibrium
one which are studied in the present article, while the domain
pinning effects can be taken into account by minor generalization
of the method used in this paper.

The paper is organized as follows. In Section \ref{sec2_model} we introduce the basic equations which we use further to evaluate the energy of the domain structure. In Section \ref{sec3_BL_barrier} we analyze the vortex penetration threshold. In Section \ref{sec4_eq_vortex_density} we give the details of calculations of the vortex density profiles. The Section \ref{sec5_eq_energy} is the central part of the paper where we compare the energies of different domain configurations and describe different possible scenarios of the first order phase transitions. In Section \ref{sec6_Conclusion} we summarize our results.

%%%%%%%%%%%%%%%%%%%%%%%%%%%%%%%%%%%%%%%%%%%%%%%%%%%%%%%%%%%%%%%%%%%%%%%%%%%%%%%%%%%%
\section{Basic equations.}\label{sec2_model}
Throughout this section we introduce the basic equations and the assumptions used in our calculations.
%Formally we divide the section into the subsections corresponding to each assumption.

\paragraph*{ Well-developed domain structure.}
As it was mentioned above we consider the emergence of superconductivity deeply in the ferromagnetic state of FS, therefore we assume that the exchange energy which depends on the absolute value $M$ of magnetization ${\bf M}$ is the largest energy in the system. This allows us to consider $M$ equal to the magnetization saturation value $M_0$.
We also assume the magnetic anisotropy to be sufficiently strong to keep the magnetization $\bf M$ oriented along the easy axis direction.
To our best knowledge these assumptions are valid for most of FS.

In these assumptions we consider a film of the ferromagnetic superconductor (FS) of thickness $2L_z$ with the easy axis of magnetization (axis~$z$) chosen perpendicular to the film.
It is useful to put the origin of this axis in the center of the FS film to have the points inside the film at $|z|<L_z$.
We put other two dimensions of the film ($L_x$ and $L_y$) to be sufficiently larger than the film thickness $L_z$.

\begin{figure}[t]
\centerline{
\includegraphics[width=0.5\linewidth]{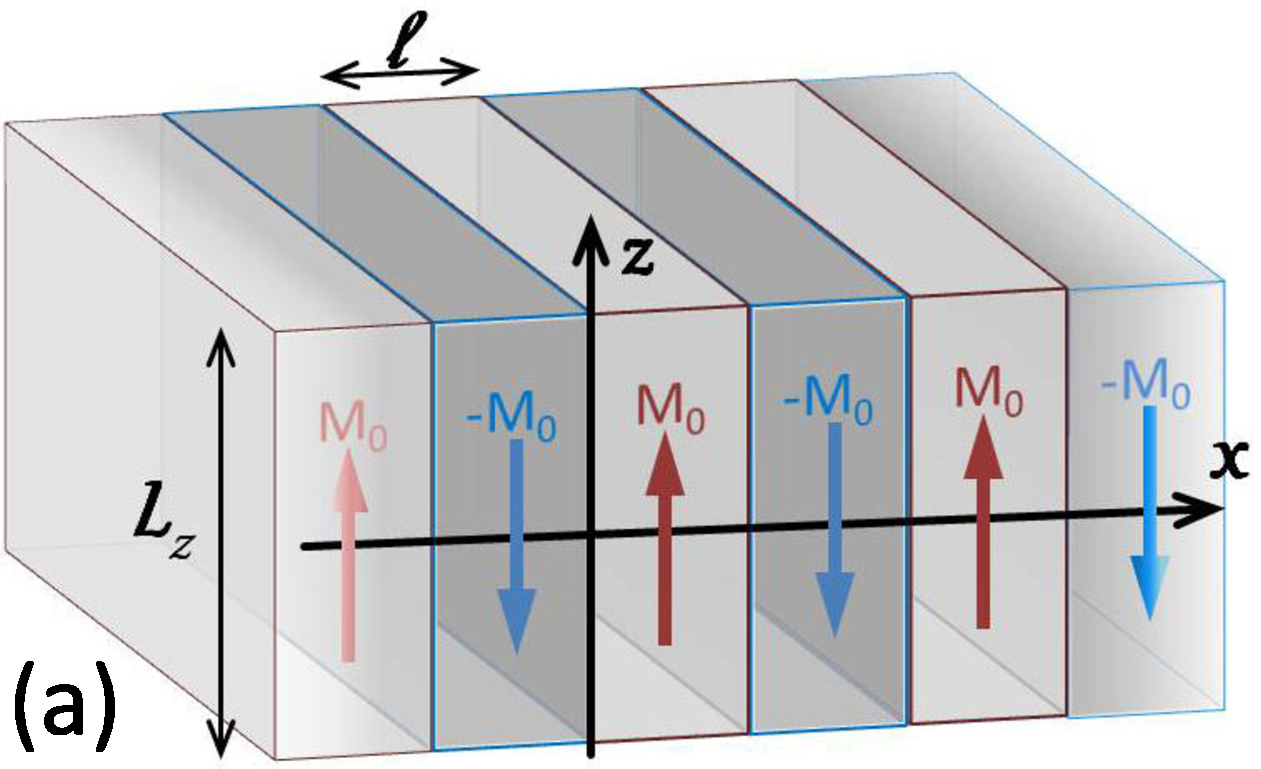}
\includegraphics[width=0.5\linewidth]{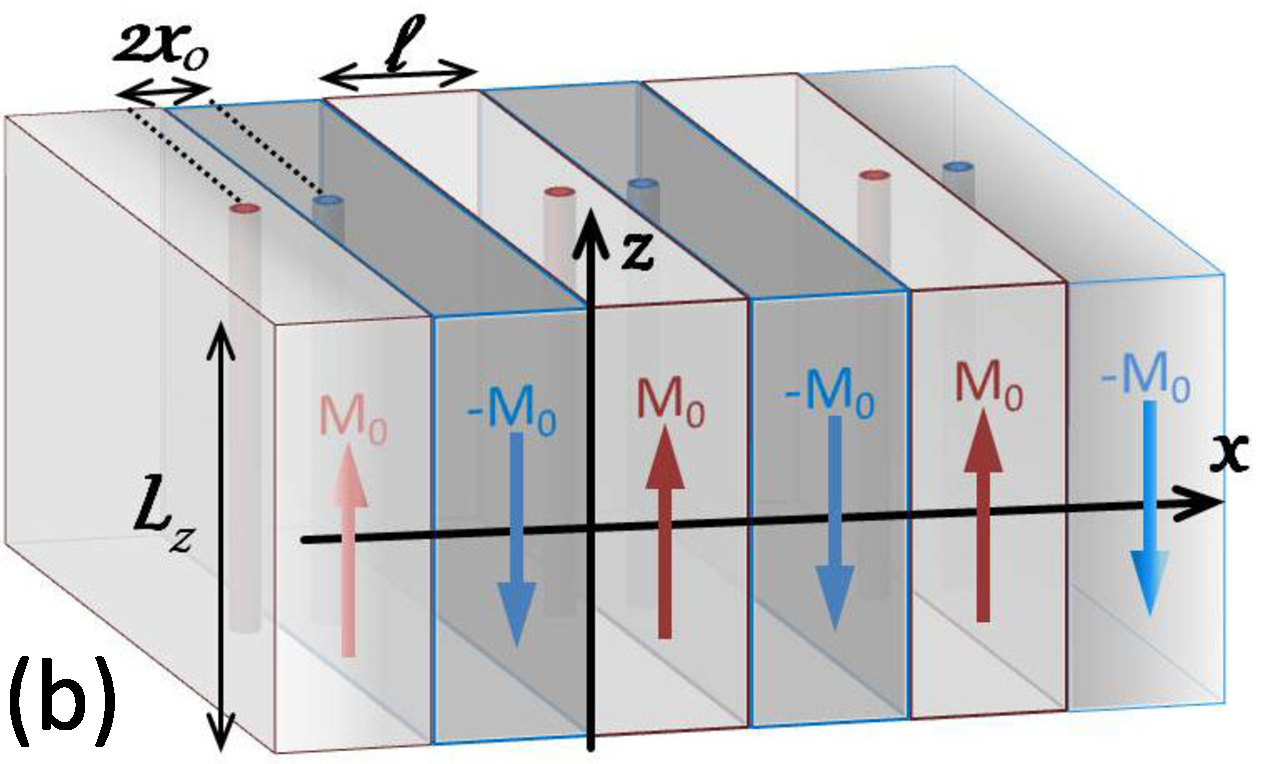}
}
\centerline{
\includegraphics[width=0.5\linewidth]{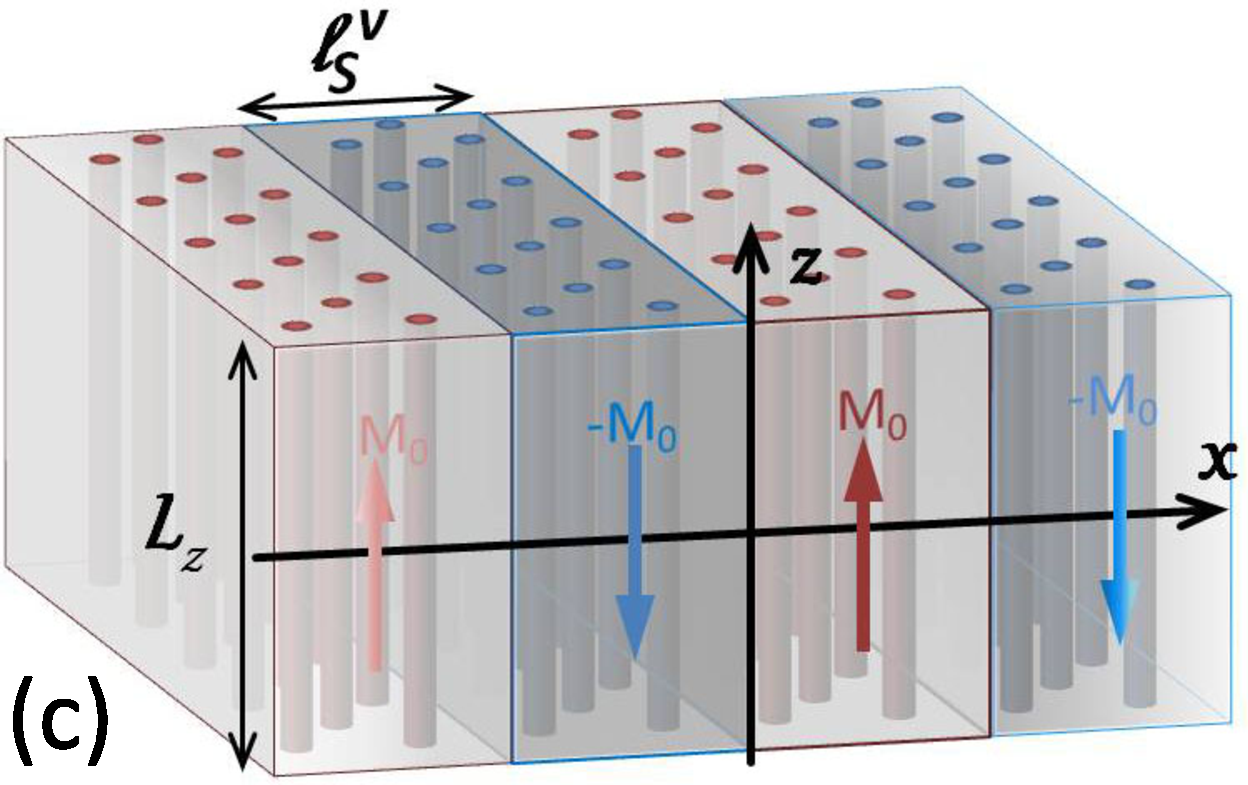}
\includegraphics[width=0.5\linewidth]{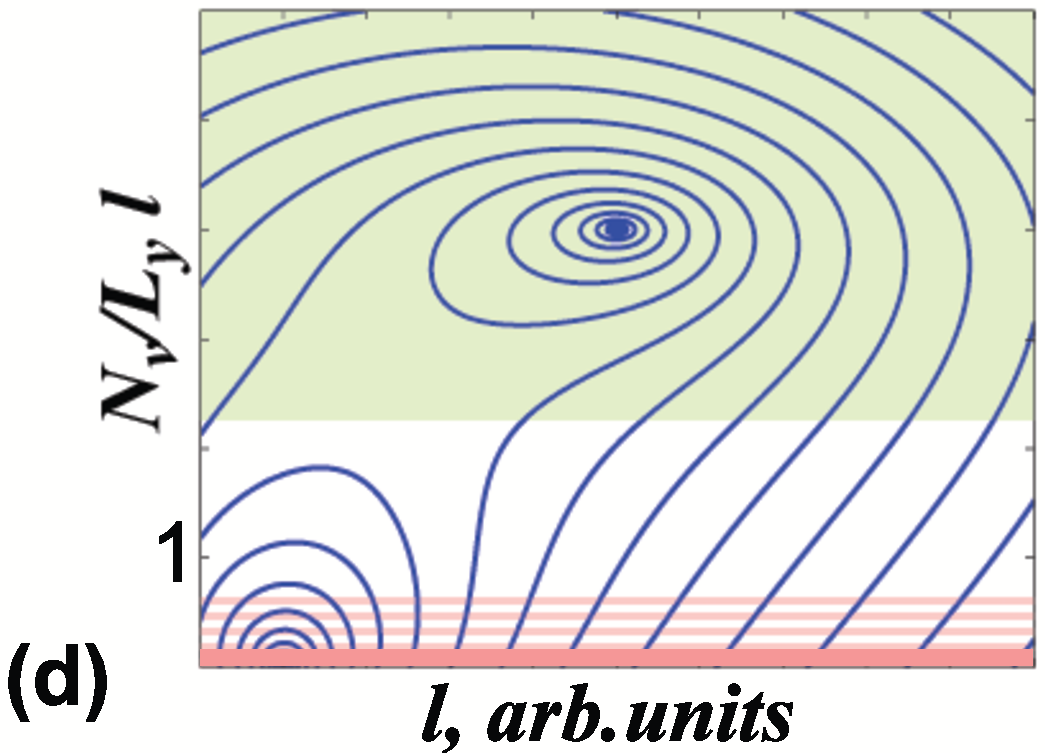}
} \caption{(a-c) The sketch of the magnetic domain structure in
the ferromagnetic superconductor of the thickness $2L_z$ with the
stripe-structure of the domains with the size $l$: (a) in the
Meissner state; (b) in the vortex state with one vortex (or
antivortex) in each domain with the distance $x_0$ between each of
them and the nearest domain wall; (c) in the vortex state with the
dense vortex lattice; (d) The sketch of the applicability of the
model at the diagram of the domain size $l$ and the vortex number
$N_V/l L_y$ in each domain. The horizontal red solid line
corresponds to the Meissner state results, the cross-hatched
region near the horizontal axis stands for calculations of the
Bean-Livingston barrier profile and the vortex penetration
threshold, and the green (shaded) region corresponds to the
validity of the dense vortex lattice results. The contour plot of
the total energy is drawn by blue curves.}
\label{Fig:SCferro_stripe-structure}
\end{figure}

Magnetic structure of the film is chosen to be the set of equal-sized domains of widths $l$ (see Fig.~\ref{Fig:SCferro_stripe-structure}(a)) with the magnetization periodically varying along the $x$ axis perpendicular to the domain walls:
\begin{equation}\label{M(r)}
{\bf M}(x) = {\bf z}_0 M_0 s(x)\theta(L_z - |z|) \ .
\end{equation}
Here periodic step function $s(x)=\sum_{m=0}^\infty 4\sin(q x)/[q l]=s(x+2l)$ with $q=\pi(2m+1)/l$ is equal to $+1$ for $0<x<l$ and to $-1$ for $l<x<2l$, while $\theta(z)$ is Heaviside theta-function: $\theta(z)=1$ for $z>0$ and $\theta(z)=0$ for $z<0$.

%%%%%%%%%%%%%%%%%%%%%%%%%%%
%%%%%%%%%%%%%%%%%%%%%%%%%%%

The equilibrium domain size (DS) $l$ of the magnetic structure in the normal state is defined by the interplay of domain wall energy (per unit area in lateral direction) $E_{DW}(l) = M_0^2{\tilde w L_z}/{l}$ inversely proportional to the DS $l$ and the magnetostatic energy contribution $E_H=\int{\bf H_0}^2dV/(8\pi S)$ of the stray magnetic fields $\bf H_0$ [\onlinecite{Kittel,LL}], which increases with the increase of the DS $l$.
Here $\tilde w$ is the effective domain wall thickness determined by the energy balance between the exchange interaction and the anisotropy cost\cite{Kittel}
and providing an upper limit of the real domain wall width\cite{Buzd_Faure} and $S=L_x L_y$ is the area of the film in the lateral direction.

The superconducting currents can essentially change the magnetic structure of the system by the additional contribution to the energy consisting of the kinetic energy of superfluid currents $\bf j$ and the energy of the current-induced magnetic fields ${\bf H-H}_0={\bf H}_M+{\bf H}_V$.
As a result the total free energy $E=E_{DW}+E_{vol}$ contains the domain-wall energy $E_{DW}$ (see above) and the volume energy $E_{vol}$:\cite{Buzd_Dao}
\begin{gather}\label{E_vol}
E_{vol}=\frac{1}{8\pi S}\int_{\Re^3}{\bf H}^2dV + \frac{2\pi\lambda^2}{c^2S}\int_{\Re^3}{\bf j}^2dV \ ,
\end{gather}
with the London penetration depth $\lambda$, light velocity in vacuum $c$, and the film area $S=L_x L_y$ in lateral dimensions.

Focusing on the vortex lattice effects we separate the volume
energy into common parts $E_{vol}=E_{MH}+E_{MV}+E_{V}$ (cf.
[\onlinecite{Buzd_Dao}]):
\begin{multline*}
E_{MH}=\frac{1}{8\pi S}\int\left[{\bf H}_0+{\bf H}_M\right]^2dV + \frac{2\pi\lambda^2}{c^2S}\int{\bf j}_M^2dV=\\=
- \frac{1}{2S}\int{\bf M}({\bf H}_0+{\bf H}_M) dV
%= 4\pi M_0^2 L_z\left[ 1-\frac{2\Lambda}{L}\tanh\left(\frac{L}{2\Lambda}\right)\right]+32\pi M_0^2 L_z \frac{1}{L^2}\sum\limits_{n=0}^\infty\frac{Q}{Q_z^4\left[1+{Q}/{Q_z} \coth Q_z\right]}
\end{multline*}
is the volume energy in the Meissner state with vortex-free superconducting currents ${\bf j}_M$ and the Meissner field ${\bf H}_M$.

The second contribution describes the interaction of the Meissner currents ${\bf j}_M$ and the vortex ones ${\bf j}_V$ :
\begin{multline}\label{E_MV}
E_{MV}=\frac{1}{4\pi S}\int{\bf H}_M{\bf H}_V dV + \frac{4\pi\lambda^2}{c^2S}\int{\bf j}_M{\bf j}_V dV=\\=
\frac{\Phi_0}{4\pi S}\int\limits_{\Re^2}\left<H_{Mz}\right>_z({\bm \rho})n({\bm \rho})d^2\rho \ ,
\end{multline}
where $\Phi_0=\pi\hbar c/e$ is the flux quantum, $\left<H_{Mz}\right>_z=\int_{-\infty}^{\infty}({\bf H}_M({\bf r})\cdot {\bf z}_0) dz$ is the $z$-component of the Meissner field averaged over the $z$-axis and the vortices with winding numbers $v_i=\pm 1$ are distributed as $n({\bm\rho})=\sum_{i=1}^{N_V}v_i\delta\left(\bm{\rho - \rho}_i\right)$ at the in-plane positions ${\bm\rho}_i=(x_i,y_i)$.

The free energy of the vortex subsystem
\begin{multline}\label{E_V}
E_{V}\cdot S =\frac{1}{8\pi}\int{\bf H}_V^2dV + \frac{2\pi\lambda^2}{c^2}\int{\bf j}_V^2dV=\\=
\frac{1}{2c}\int{\bf j}_V\left[\frac{4\pi\lambda^2}{c}{\bf j}_V+{\bf A}_V\right]dV = \varepsilon_V^0 N_V + U_{vv} \ ,
\end{multline}
contains the self energy $\varepsilon_V^0$ of each of $N_V$ vortices/antivortices in the FS sample and the vortex interaction energy
\begin{gather}\label{U_vv_sum}
U_{vv} = \frac{1}{2}\sum\limits_{i\neq j}v_i v_j V_0\left(|{\bm \rho}_i-{\bm \rho}_j|\right) \ ,
\end{gather}
determined by the potential of the interaction $V_0(R)$ between two isolated vortices situated at the distance $R$ from each other.

\paragraph*{London approximation.}
The further assumption used in our model is that the superconducting coherence length $\xi(T)$ is small compared with with DS and penetration depth $\xi(T)\ll\lambda(T),l(T)$ for all temperature values $T<T_c$ to make the London-type equations valid:
\begin{gather}\label{London_eq_Meissner}
{\rm rot rot}{\bf A}_\lambda ({\bf r})= 4\pi{\rm rot}{\bf M}({\bf r})-\frac{1}{\lambda^2}\theta(L_z - |z|){\bf A}_\lambda ({\bf r}) %\equiv 4\pi{\rm rot}{\bf M}({\bf r})+\frac{4\pi}{c} {\bf j}_M ({\bf r})
\ ,
\end{gather}
with the last term corresponding to the Meissner-induced screening current as follows ${\bf j}_M ({\bf r})=-\theta(L_z - |z|)\cdot{\bf A}_\lambda ({\bf r})c/(4\pi\lambda^2)$.
The latter equation is written for the vector-potential ${\bf A}_\lambda({\bf r})$ and the superconducting currents ${\bf j}_M$ in the Meissner state [${\bf H}_0+{\bf H}_M+4\pi{\bf M}={\rm rot}{\bf A}_\lambda({\bf r})$ with the magnetization given by \eqref{M(r)}] and reduces to the normal state one ${\bf H}_0=\rm rot {\bf A}_\infty ({\bf r})$ for $\lambda\to\infty$.

The solution of the London equation \eqref{London_eq_Meissner} for vortex-independent part of vector-potential ${\bf A}_\lambda({\bf r})$ takes the standard form of expansion to the Fourier series
$${\bf A}_\lambda({\bf r})=- {\bf y}_0 \frac{16\pi M_0}{l} \sum_{m=0}^\infty a_\lambda(q,z)\cos(q x) \ ,$$
with
$$a_\lambda(q,z)=\left\{\frac{1}{q_z^2}-\frac{q L_z\cosh(q_z z)}{q_z C_q\sinh (q_z L_z)}\ , |z|<L_z\atop
\frac{L_z}{C_q}\exp[-q(|z|-L_z)]\ , |z|>L_z\right. \ ,$$
\begin{gather}\label{C_q}
C_q = L_z q_z^2\left[1+\frac{q\coth(q_z L_z)}{q_z}\right] \ ,
\end{gather}
%$$a_\lambda(q,z)=\left\{1/q_z^2-q \cosh(q_z z)/[q_z C_q\sinh (q_z L_z)]\ , |z|<L_z\atop
%\exp[-q(|z|-L_z)]/C_q\phantom{1-q \cosh(q_z z)}\ , |z|>L_z\right. \ ,$$
$q=\pi(2m+1)/l$, $q_z^2=q^2+\lambda^{-2}$, and integer $m$.
As a result the vortex-independent part of the volume energy can be written in the form
\begin{multline}\label{E_MH}
E_{MH}= - \frac{1}{2S}\int M({\bf r})\left[\partial_x A_\lambda({\bf r})-4\pi M({\bf r})\right] dV =\\=
 \frac{E_{M}}{l^2}\sum\limits_{n=0}^\infty\left[\frac{1}{q^2}-\frac{1}{q_z^2}+\frac{q}{q_z^2 C_q }\right]
\ ,
\end{multline}
with the characteristic energy scale (per unit area) $E_M=32\pi
M_0^2 L_z$. Obviously the expression for $E_{MH}$ coincides with
the volume energy of the Meissner state derived in
[\onlinecite{Buzd_Dao}].

Analogously we have the following expression for the averaged Meissner field $\left<H_{Mz}\right>_z$, which enters the expression \eqref{E_MV} for energy $E_{MV}$:
\begin{multline}\label{h(rho)}
\left<H_{Mz}\right>_z=\int_{-\infty}^{\infty}\partial_x\left[A_\lambda({\bf r})-A_\infty({\bf r})\right] dz =\\=
-\frac{8\pi M_0 L_z}{\lambda^2}\sum\limits_{n=0}^\infty \frac{4\sin(q x)}{q l} f_q\ ,
\end{multline}
with $f_q = \left[1-{q}/{C_q }\right]/{q_z^2}$. Here the subscript
$\infty$  in $A_\infty$ means the limit of a large London
penetration depth $\lambda\to\infty$ corresponding to its normal
value. We wrote the sum \eqref{h(rho)} in the similar form to
\eqref{M(r)} to emphasize the analogy between
$\left<H_{Mz}\right>_z$ and the periodic profile of the
magnetization $M(x)=M_0\sum_{m=0}^\infty 4\sin(q x)/[q l]$.

For calculating the vortex energy in the form of r.h.s. of Eq.~\eqref{E_V} we use linearity of London equations and put the vortex-induced vector-potential ${\bf A}_V ({\bf r})$ [${\bf H}_V ({\bf r})=\rm rot {\bf A}_V ({\bf r})$] in the form of the superposition of the screening current contributions near each of the vortices:
\begin{gather}
{\bf A}_V ({\bf r}) = \sum\limits_{i=1}^{N_V}v_i{\bf A}_V^{(0)}\left(\bm{ r - \rho_i}\right) \ .
\end{gather}
with ${\bf A}_V^{(0)}({\bf r})$ satisfying the London equation:
\begin{multline}
{\rm rot rot}{\bf A}_V^{(0)} ({\bf r})= -\frac{1}{\lambda^2}\theta(L_z - |z|)\left({\bf A}_V^{(0)} ({\bf r})-\frac{\Phi_0{\bf e}_{\varphi}}{2\pi|{\bm\rho}|}\right) \equiv\\ \equiv \frac{4\pi}{c} {\bf j}_V^{(0)} ({\bf r}) \ .
\end{multline}
for an isolated Abrikosov vortex with positive vorticity $v_0=1$ situated at the origin $\bm\rho_0=0$.
Here ${\bf e}_{\varphi} = [{\bf z}_{0}\times {\bm \rho}]/\rho$ is the unit vector along azimuthal angle $\varphi$.

The latter equation has the well-known solution for an arbitrary ratio $\lambda/L_z$ (see, e.g., [\onlinecite{Brandt,Meln_Buzdin_pancake}]) within the London gauge and the cylindrical symmetry ${\bf A}_V^{(0)} = {\bf e}_{\varphi} A_V^{(0)}(\rho,z)$.
As a result the expression for the Fourier component of the vortex-vortex interaction term $V_{\bf G}=(2\pi L_z^2)^{-1}\int V_0(R)e^{-i{\bf G R}}d^2 R$
in the superconducting film with arbitrary thickness $L_z$ takes the form
\begin{gather}\label{V_G}
V_G = \frac{\left({\Phi_0}/{2\pi\lambda}\right)^2 L_z^{-1}}{G^2+\lambda^{-2}}\left[1+\frac{1}{\lambda^{2}G C_G}\right] \ ,
\end{gather}
with $C_G$ derived from \eqref{C_q} by substituting $G$ instead of $q$, % = L_z G_z^2\left[1+G/G_z\coth(G_z L_z)\right]$
 and $G_z^2 = \lambda^{-2}+G^2$.
Strictly speaking this expression works well for $|{\bf G}|\lesssim \xi^{-1}$, because the interaction energy \eqref{U_vv_sum} doesn't contain the self energies $\varepsilon_V^0\cdot N_V$ of the vortices.
In the high $G$ limit Eq.~\eqref{V_G} should be modified in such a way to have $V_0(|\bm\rho_i-\bm\rho_j|<\xi)=0$.
The self energy of each vortex $\varepsilon_V^0$ can be written using this Fourier component $V_G$ in a such way
\begin{multline*}%\label{E_V^0_result}
\varepsilon_V^0 = \frac{L_z^2}{4\pi}\int\limits_{\Re^2}V_G d^2{\bf G}\approx \frac{L_z^2}{2}\int\limits_{0}^{1/\xi}V_G G dG\approx \\ \approx 2L_z\left(\frac{\Phi_0}{4\pi\lambda}\right)^2 \left\{\ln\tilde\kappa+\frac{1}{2}+O\left[\min\left(\frac{\lambda}{L_z},\frac{L_z^{2}}{\lambda^{2}}\right)\right]\right\} \ ,
\end{multline*}
with the Landau Ginzburg parameter $\tilde\kappa=\lambda^*/\xi$ and the effective penetration depth $\lambda^*=\max(\lambda,\lambda^2/L_z)$ of the magnetic field.
We have to cut off logarithmic divergence in the integration for the vortex self energy in high $G$ limit of order of inverse vortex core size $\xi^{-1}$.

\paragraph*{ Dense vortex lattice and the correlation between magnetization sign and the vorticity.}
Further analysis is based on the assumption that the vortex lattice is dense at the lengths of order of the domain size. This assumption allows us to consider the continuous model of the vortex distribution function $n({\bm\rho})$.
The vortex-vortex interaction energy $U_{vv}$ in terms of the vortex density takes the form\cite{Pokrovsky}:
\begin{gather}\label{U_vv}
U_{vv} = \frac{1}{2}\iint%_{|\bm\rho-\bm\rho'|>\xi}
V_0(|{\bm\rho- \bm\rho'}|)n({\bm\rho})n({\bm\rho}')d^2{\bm\rho}d^2{\bm\rho}' \ ,
\end{gather}
where the condition $\bm\rho\neq\bm\rho'$ included into Eq.~\eqref{U_vv_sum} is taken into account by the choice $V_0(|\bm\rho_i-\bm\rho_j|<\xi)=0$ discussed above.
%by the $|\bm\rho-\bm\rho'|>\xi$ in the continuous consideration of the vortex density.
%
%\texttt{TO DO!!!}
%\begin{multline*}
%\tilde V_{\bf G}=\frac{1}{2\pi L_z^2}\int\limits_{R>\xi} V_0(R)e^{-i{\bf G R}}d^2 R=V_{\bf G}-\\-\frac{1}{2\pi L_z^2}\int\limits_{R<\xi} V_0(R)e^{-i{\bf G R}}d^2 R
%\approx V_{\bf G}-\frac{\pi\xi^2}{2 \pi L_z^2}V_0(0)=V_{\bf G}-\frac{\xi^2}{L_z^2}\varepsilon_V^0
%\end{multline*}
%
%In the case of the rare vortex lattice, when the vortex distribution function $n({\bf r})$ is written in the form of delta-functions
%one should know that Eq.~\eqref{U_vv} takes into account also the self-vortex energy \eqref{E_V} comparing to Eq.~\eqref{U_vv_sum}.
%As a result we should subtract the single vortex energy term from the vortex free energy \eqref{E_V} when we put the vortex distribution function in the form of delta-functions into Eq.~\eqref{U_vv}.

The number of the vortices $N_V$ can be easily written as the integral of the vortex-antivortex distribution function $n({\bm\rho})$ \cite{Pokrovsky}
\begin{gather}\label{N_V}
N_V = \int_{\Re^2}n({\bm\rho})s({\bm\rho})d^2{\bm\rho}
\end{gather}
in the assumption that in the equilibrium vortex-antivortex density corresponds to the situation when the vortices located in the domains with positive magnetization sign $s(x)={M_z(x)}/{M_0}$ and antivortices - with negative one, i.e. vorticity is equal to the sign of the magnetization $v_i=s(x_i)$.

%One can see that \eqref{U_vv} coincides with Eq.~(4) in [\onlinecite{Pokrovsky}]. But the single vortex energy in \eqref{E_V} differs from (3) in [\onlinecite{Pokrovsky}] by $-M_0\Phi_0 N_V/2$, where it was included into $E_{MV}$ or taken into account twice (see Eqs.~(3, 5) in [\onlinecite{Pokrovsky}]) by mistake.

Using the Fourier transformation $n({\bm\rho})=\sum_{\bf G}n_{\bf G}e^{i{\bf G\bm\rho}}$ and substituting
$s(x)=\sum_{m=0}^\infty 4\sin(q x)/[q l]$, \eqref{h(rho)}, and \eqref{V_G} to (\ref{E_MV}, \ref{E_V})
one can obtain the following expression for the vortex-dependent part of the energy $E_{nV} =E_{V}+E_{MV}$:
\begin{multline}\label{E_nV_n_G}
E_{nV}= \sum\limits_{\bf G}\left[\left(\varepsilon_V^0-\frac{2\Phi_0 M_0 L_z}{\lambda^2} f_{q}\right)s_{\bf G}n_{-\bf G}+\right.\\ \left.+\pi L_z^2 V_{\bf G} n_{\bf G}n_{-\bf G}\right] \ ,
\end{multline}
with $s_{\bf G}=-2i\delta_{G_y,0}\left(\delta_{G_x,q}-\delta_{G_x,-q}\right)/q l$.
Here we assume that the sample has the large sizes in lateral dimensions $L_x, L_y\gg \lambda,l,L_z$.
Note that the modification of expression \eqref{V_G} for $V_{\bf G}$ at large $|{\bf G}|$ discussed above plays the role only for $n_{\bf G}$ with rather wide spectrum, i.e. in the case when the vortex distribution function $n({\bf r})$ is written in the form of delta-functions.
However, even in this case one can also use the expression \eqref{V_G} for $|{\bf G}|\sim\xi^{-1}$ if one additionally subtract a certain constant term from $V_{\bf G}$ to get $V_0(|\bm\rho_i-\bm\rho_j|<\xi)=0$.
Occasionally, the absolute value of this additional term is equal to the single energy term of $N_V$ vortices, therefore for the delta-functional vortex density one can use \eqref{E_nV_n_G} with $V_{\bf G}$ from \eqref{V_G} if one effectively threw the single energy term away from vortex-dependent energy.

As we will see below the parameter
\begin{gather}\label{l_alpha}
l_v=(\Phi_0 \ln\tilde\kappa/2\pi^2 M_0)^{1/2}\sim (\Phi_0/M_0)^{1/2}
\end{gather}
which is the typical intervortex distance for the magnetic fields of order of magnetization amplitude $H\sim M_0$ relative to the other characteristic lengths determines the vortex penetration threshold and play an important role in the equilibrium DS.

The basic idea of the rest part of the paper is schematically
sketched in Fig.~\ref{Fig:SCferro_stripe-structure}(d). In rather
general case of the parameters the shown contour plot of the total
energy minimized over the vortex distribution has two local minima
versus the domain size $l$ and the number of the vortices in each
domain. One of these minima is in the Meissner state, which
corresponds to the $l$-axis, and another one is somewhere at the
finite vortex density. Using the described model one can calculate
the energy profile in the Meissner state. Substituting the vortex
distribution with the one vortex (or antovortex) in each domain
(see Fig.~\ref{Fig:SCferro_stripe-structure}(b)) to the total
energy one can find the vortex penetration threshold with the
fixed domain size and check if the Meissner state realizes at least a local energy minimum for the certain parameters. The total energy
calculations with the dense vortex lattice can give us the most
favorable vortex space distribution and the total energy minimum
which can be compared with the value in the Meissner state.

As a result, in rather general situation we can consider the phase transitions between Meissner and the vortex states with decreasing temperature as follows.
For the temperatures larger than SC critical temperature the domain structure of the FS is equivalent to the normal ferromagnetic one.
Lowering of the temperature leads to the emergence of the superconductivity accompanied by the spontaneous vortex penetration into the FS. The vortices prevent the domain structure to shrink sufficiently (see Fig.~\ref{Fig:SCferro_stripe-structure}(c)).
At some lower temperatures there is the critical temperature where the vortex state is not stable anymore and the ground state corresponds to the Meissner state with rather different domain size (see Fig.~\ref{Fig:SCferro_stripe-structure}(a)).
As we will see below this phase transition of abrupt changing of the vortex number and the domain size is the first-order phase transition.

%====================
%
%To simplify the expressions for the energy it is convenient to normalize all the lengths
%to the half of  FS film thickness $L_z$ (similar to [\onlinecite{Buzd_Dao}]): the domain size $L=l/L_z$, the London $\Lambda=\lambda/L_z$ and effective $\Lambda^*=\max(\Lambda,\Lambda^2)$ penetration depths, and the effective domain wall width $W=\tilde w/L_z$.
%
%To further simplify the expressions we introduce a normalized energy $E$ as
%$E=32\pi M_0^2 L_z \cdotE$ and express the terms $E_{DW}$, $E_{MH}$, and $E_{nV}$  in similar notations:
%\begin{gather}\label{E_MH}
%E_{DW}=\frac{W}{32\pi L}\ , \quad
%%\end{gather}
%%\begin{gather}
%E_{MH}= \frac{1}{8}\left[ 1-\frac{2\Lambda}{L}\tanh\left(\frac{L}{2\Lambda}\right)\right]+\frac{1}{L^2}
%\sum\limits_{n=0}^\infty\frac{Q}{Q_z^3\left[Q_z + Q\coth Q_z\right]}\ ,
%\end{gather}
%\begin{gather}\label{E_nV}
%E_{nV}= \frac{2}{\pi^2\Lambda^2}\left[\sum\limits_{m}\frac{\left(\alpha-f_Q\right)\left(\rho_Q-\rho_{-Q}\right)i}{2m+1}+\sum\limits_{\bf G}u_{|{\bf G}|}\rho_{\bf G}\rho_{-\bf G}\right] \ ,
%\end{gather}
%where $\rho_a = \Phi_0/(16 M_0) n_a$ is dimensionless Fourier component of the vortex density and $\alpha = \Lambda^2\varepsilon_V^0/(2\Phi_0 M_0 L_z) = \Phi_0/M_0 (4\pi L_z)^{-2} \ln\tilde\kappa$. Here we introduce the exact definition of the intervortex length $l_v$ as $L_{\alpha}=l_v}/L_z=(8\alpha)^{1/2}$.

%%%%%%%%%%%%%%%%%%%%%%%%%%%%%%%%%%%%%%%%%%%%%%%%%%%%%%%%%%%%%%%%%%%%%%%%%%%%%%%%%%%%
\section{Vortex penetration threshold. Bean-Livingston barrier profile.}\label{sec3_BL_barrier}
The most probable scenario of (anti)vortex penetration into the sample in the geometry given by Fig.~\ref{Fig:SCferro_stripe-structure}(a) is implemented by the creation of the vortex-antivortex pairs at the domain wall.
To calculate the profile of the Bean-Livingston barrier and find the minimum value of the magnetization which is enough to make vortex state energetically favorable for certain domain size one should substitute the vortex density into the vortex-dependent energy $E_{nV}$  in the form of the periodic lattice of the vortex-antivortex pairs with the distances from the nearest domain wall equal to certain distance $x_0$ (see Fig.~\ref{Fig:SCferro_stripe-structure}(b) for details):
\begin{multline*}
n({\bm\rho})=\delta_\xi(y)\sum_{m'=-\infty}^{\infty}\left[\delta_\xi(x-x_{m'}^+)-\delta_\xi(x+x_{m'}^-)\right]=\\=
\sum_{k}\sum_{|G_y|\lesssim\xi^{-1}}n_k\sin\left(k x\right) e^{i G_y y} \ ,
\end{multline*}
where $x_{m'}^\pm=\pm x_0 + 2l m'$ is the position of the $m'$th (anti)vortex, $\delta_\xi(x)$ is the physical delta-function with the length scale $\xi$, $n_k={2}\sin\left(k x_0\right)/({l L_y})$ is the Fourier components of the $n({\bm\rho})$ for momentum projections $k=\pi (m+1)/l$ to the axis $Ox$ perpendicular to the domain walls and $G_y=2\pi m_y/L_y$ to the axis $Oy$ parallel to them, with integer $m$ and $m_y$. Here and further we assume that the sample size along $y$-axis $L_y$ is much larger than the domain size $L_y\gg l$.

After modification of $V_{\bf G}$ by subtracting of the constant term equal to the self energies of $N_V$ vortices [see discussion after \eqref{E_nV_n_G}] we obtain the profile of Bean-Livingston energy barrier depending on the distance $x_0$:
\begin{multline}\label{E_nV_th}
E^{th}_{nV}(x_0) = \frac{E_0}{\lambda^2}\sum\limits_{m=0}^{2 N_\xi} \frac{\Phi_0}{16\pi M_0 l}\sin^2\left[k x_0\right]\left<u\right>(k)-\\-
\frac{E_0}{\lambda^2}\sum\limits_{m=0}^{N_\xi}\frac{f_{q}}{2m+1}\sin\left[q x_0\right] \ ,
\end{multline}
with the function $f_q = \left[1-{q}/{C_q }\right]/q_z^2$, $C_q$ given by \eqref{C_q}, the typical barrier amplitude $E_0 = {8\Phi_0 M_0 L_z}/{(\pi l L_y)}$ (per unit area in lateral dimension), momenta $k=\pi (m+1)/l$, $q=\pi(2m+1)/l$, and the normalized vortex-vortex interaction term \eqref{V_G} $u_q=G_z^{-2}\left[1+(\lambda^{2}G C_G)^{-1}\right]$
\begin{multline*}
\left<u\right>(k)=\sum_{|G_y|<\xi^{-1}} \frac{{\pi}/{L_y}}{G^2+\lambda^{-2}}\left[1+\frac{1}{\lambda^{2}G C_G}\right]\approx\\ \approx
\int_0^{\xi^{-1}}\frac{dG_y}{G^2+\lambda^{-2}}\left[1+\frac{1}{\lambda^{2}G C_G}\right]  \ ,
\end{multline*}
averaged over $y$-axis projection $G_y$ of the momentum ${\bf G}=(k,G_y)$. The divergent terms should be cut off at the momenta of order of the inverse vortex core size $G_y, k, q\sim\xi^{-1}$, i.e. $m\sim N_\xi=l/(2\pi\xi)$. Here we used the integration instead of summation over $G_y$, due to large size $L_y$ of the sample along $y$-axis.

%%%%%%%%%%%%%%%%%%%%%%%%%%%%%%%%%%%%%%%%%%%%%%%%%%%%%%%%%%%%%%%%%%%%
\begin{figure}[t]
\includegraphics[width=0.4\textwidth]{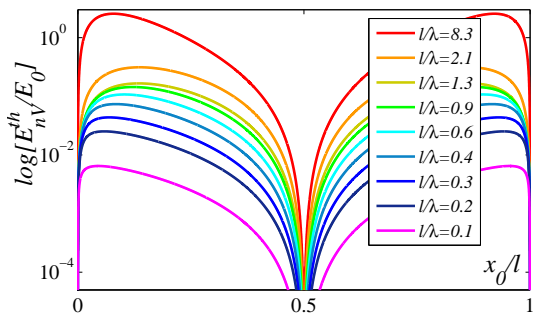}
\caption{The energy profile $E_{nV}^{th}(x_0)$ of the vortex penetration barrier normalized to $E_0$ in logarithmic scale for decreasing ratio $l/\lambda$ from top to bottom.
All the plots are shown for the threshold magnetization values for the certain domain size $l$ and for $\tilde w/L_z=10^{-4}$.}
 \label{Fig:SCferro_BL_barrier}
\end{figure}
%%%%%%%%%%%%%%%%%%%%%%%%%%%%%%%%%%%%%%%%%%%%%%%%%%%%%%%%%%%%%%%%%%%%

While the vortex-dependent threshold energy \eqref{E_nV_th} has only non-negative values, the vortex state can't be stable for the certain domain size.
Increasing magnetization amplitude $M_0$ one can effectively suppress the contribution of the first term in \eqref{E_nV_th} relatively to the second one.
At the certain magnetization value the energy profile turns to negative value at some point and the vortex state becomes more favorable than the Meissner one for the fixed domain size.
The typical Bean-Livingston barrier profiles at the threshold magnetization values for different ratios $l/\lambda$ of the domain size $l$ and London penetration depth $\lambda$ are shown at Fig.~\ref{Fig:SCferro_BL_barrier}.
The domain size $l$ for all the plots is chosen to be the one $l_S$ which minimizes the Meissner energy of FS (see Sec.~\ref{sec5_eq_energy} for details).

Note that for whole range of parameters the most energetically favorable place for vortices is at the center of each domain $x_0=l/2$.
Considering the energy value at this point:
\begin{gather}\label{E_nV_th_1/2}
E^{th}_{nV}(l/2) = \frac{E_0}{\lambda^2}\sum\limits_{m=0}^{N_\xi}\left[ \frac{\Phi_0\cdot \left<u\right>(q)}{16\pi M_0 l}
-\frac{(-1)^m f_{q}}{2m+1}\right] \ ,
\end{gather}
and using the inequality $q<C_q<\infty$ (i.e. $q_z^{-2}<u_q<q^{-2}$ and $\pi/2q_z<\left<u\right>(q)<\pi/2q$) one can estimate the bounds of the sum $\sum_{m=0}^{N_\xi}\left<u\right>(q)$  as follows
 $$\frac{l}{4}\ln \kappa_m\lesssim\sum_{m=0}^{N_\xi}\left<u\right>(q)\lesssim\frac{l}{4}\ln \left(\frac{l}{\xi}\right) \ ,$$
with $\kappa_m = \min(l,\lambda^*)/\xi$.
The upper bound can be reached for small domain sizes $l\ll\lambda^*$, while the lower one~--~in the opposite limit $l\gg\lambda^*$.
The second term in \eqref{E_nV_th_1/2} converges quickly enough to extend the sum to infinity.

As a result the threshold value of the magnetization takes the form
\begin{gather}\label{M_th}
4\pi M_{th}=\frac{\Phi_0 \ln\kappa_m}{16 }\left(\sum\limits_{m=0}^{\infty}\frac{(-1)^m f_{q}}{2m+1}\right)^{-1} \ ,
\end{gather}
with $f_q = \left[1-{q}/{C_q }\right]/{q_z^2}$, and $C_q$ given by \eqref{C_q}.

The limiting cases of the latter expression give the standard results mentioned, e.g., in [\onlinecite{Buzd_Dao}]:

(i) For rather high temperatures when screening effects are not sufficient $l\ll\lambda^*(T)$ Eq.~\eqref{M_th} reduces to %$f_Q=(e^{-2Q}-1+2Q)/(2Q^3)$
\begin{gather}\label{alph_th_L<<Lmd}
M_{th}=\frac{\Phi_0 \ln(l/\xi)}{2\pi^2 l^2}\left[1-\frac{l}{\pi^4 L_z}F\left(\frac{L_z}{l}\right)\right]^{-1} \ ,
\end{gather}
with
$F(z)=L(\tfrac{1}{2},4,\tfrac{1}{2})-L(2 i z + \tfrac{1}{2},4,\tfrac{1}{2})$ and the Lerch zeta function
$L\left(y,k,\delta\right) = \sum_{m=0}^{\infty}{\exp(2\pi i y)}/{(m+\delta)^k}$. In this case magnetization threshold both for thick films $ L_z\gg l$ [see $M_{th*}(l)$ in \eqref{M_th_L_z,lmd_gg_l}] and for thin ones $L_z\ll l$ [see $M_{th}^*(l)$ in \eqref{M_th_L_z_ll_l_ll_lmd}] is larger than the lower critical field $H_{c1}=\Phi_0\ln\tilde\kappa/(4\pi\lambda^2)$ in the bulk superconductor:
%\begin{gather}
\begin{subequations}
\begin{align}
\label{M_th_L_z,lmd_gg_l}
M_{th*}(l)&=\frac{\Phi_0 \ln(l/\xi)}{2\pi^2 l^2} \ , \quad L_z\gg l \ , \\
%\end{gather}
%\begin{gather}
\label{M_th_L_z_ll_l_ll_lmd}
M_{th}^*(l)&=\frac{\Phi_0 \ln(l/\xi)}{64 G l L_z}  \ , \quad  L_z\ll l\ \ .
%\end{gather}
\end{align}
\end{subequations}
Here $G\approx0.915966$ is Catalan's constant. These limiting cases coincide with (39, 41) in [\onlinecite{Buzd_Dao}].

(ii) In the opposite case of rather good screening of magnetic field $l\gg\lambda^*$: $f_q\approx\lambda^2$ and one can easily prove that the standard result restores
\begin{gather}\label{M_th_lmd_ll_l}
4\pi M_{th}^\lambda=H_{c1}
\end{gather}

One can easily check that in terms of the typical intervortex distance $l_v\sim(\Phi_0/M_0)^{1/2}$ for $H\sim M_0$ the vortex penetration threshold corresponds to $l_v=l_{v*}^{th}\simeq l$ for $l\ll L_z,\lambda^*$, $l_v=l_v^{th*}\simeq (4/\pi)\sqrt{2 G l L_z}$ for $L_z\ll l\ll\lambda^*$ and $l_v=l_{v,\lambda}^{th}=2\lambda\sqrt{2}$ for $l\gg \lambda^*$.
Here we neglect the difference between $\ln\tilde\kappa$ and $\ln(l/\xi)$ for simplicity.

The results derived in this section allows us to check if the Meissner state is stable to rather small fluctuations of the vortex number for the whole range of parameters.
We can also calculate the amplitude of the energy barrier for vortex penetration to make some estimation of the weak oscillating magnetic field amplitude used in [\onlinecite{dynamical_pinning_SF-1,Meln_Buzdin,dynamical_pinning_SF-2}] for equilibration of the ferromagnetic and vortex subsystems in FS.

%%%%%%%%%%%%%%%%%%%%%%%%%%%%%%%%%%%%%%%%%%%%%%%%%%%%%%%%%%%%%%%%%%%%%%%%%%%%%%%%%%%%
\section{Equilibrium vortex density distribution.}\label{sec4_eq_vortex_density}
The next two sections are devoted to the effects of the dense vortex lattice.
In this section we calculate the equilibrium distribution $n_{eq}(x)$ of the vortex density which minimizes the vortex-dependent part~\eqref{E_nV_n_G} of energy
and consider the corresponding minimum value $E_{vol}$ of the volume energy \eqref{E_vol} for the general case of parameters.
In the limiting cases we will demonstrate the difference in vortex distributions originated from the strong (weak) intervortex interaction in the thin (thick) films.

Minimizing Eq.~\eqref{E_nV_n_G} for $M_0>M_{th}(l)$ one can obtain the equilibrium vortex distribution in the form $n_{eq}(x)=\sum_{m=0}^\infty n_{q}^{eq}\sin(q x)$, with $q=\pi(2m+1)/l$ and
\begin{gather}
n_{q}^{eq}=\frac{16 M_0}{\Phi_0}\frac{f_q-l_v^2/8}{(2m+1)u_{q}} \ ,
\end{gather}
with $l_v$ given by \eqref{l_alpha}.

The vortex-dependent part $E_{nV}$ of the energy, minimized over the vortex density, takes the form
\begin{gather}\label{E_nV_min}
E_{nV} = -\frac{E_M}{\pi^2\lambda^2}\sum\limits_{m=0}^\infty{\frac{\left[f_q - l_v^2/8\right]^2}{(2m+1)^2 u_q}} \ ,
\end{gather}
with $E_M=32\pi M_0^2 L_z$.

Using the definitions of the functions
\begin{gather}\label{u_Q_f_Q}
u_q = \frac{1}{q_z^2}\left[1+\frac{1}{\lambda^{2}q C_q}\right] \ ,\quad f_q = \frac{1}{q_z^2}\left[1-\frac{q}{C_q }\right] \ ,
\end{gather}
with $C_q$ given by \eqref{C_q} one can obtain the following expressions for their ratios:
\begin{subequations}
\begin{align}
\label{u_Q^-1}\frac{\lambda^2}{u_q} =1+\lambda^2q^2-A_q%\frac{q_z^2}{\lambda^{-2}+q C_q}
 \ ,\quad
\frac{f_q}{u_q} =1-A_q%\frac{q_z^2}{\lambda^{-2}+q C_Q}
 \ ,\\
\frac{f_q^2}{\lambda^2 u_q} =1-\frac{q^2}{q_z^2}+\frac{q^3}{q_z^2 C_q}-A_q%\frac{q_z^2}{\lambda^{-2}+q C_q}
\ ,
\end{align}
\end{subequations}
with $A_q={q_z^2}/{(\lambda^{-2}+q C_q)}$.

The term $\lambda^2 q^2$ in \eqref{u_Q^-1} leads to divergence in both series for the vortex density and the energy \eqref{E_nV_min}. The sum of this term in both cases is proportional to the number of summands.
In the continuous approximation of the vortex density this divergence should be cut at the momenta $q=\pi(2m+1)/l$ equal to inverse intervortex distances $q_v\sim n(x)^{1/2}$ where this approximation stops working.
On the other hand, the intervortex distance $q_v$ should be small compared to the domain size for the dense vortex lattice $q_v l\gg 1$. This restricts our consideration to the values of magnetization far from the threshold value of the vortex penetration $M_0\gg M_{th}(l)$, i.e. $l_v\ll l_v^{th}$.
As we will see below the cutting of Fourier series works very well even till the threshold for nearly all cases, except the case of rather thin samples $L_z\ll \tilde w$ and large values of $\lambda\gg l$.

Nevertheless we can neglect the $l_v$-dependent terms for estimating the typical inverse intervortex distance $q_v$ and approximate it with the square root of the average vortex density:
$$q_v^2=\frac{1}{l}\int\limits_0^{l}n(x)dx\approx\frac{32 M_0}{\pi\Phi_0}\sum\limits_{m=0}^{\infty}\frac{f_{q}}{(2m+1)^2u_{q}} \ .$$
In other words the maximal harmonics number $N$ in this approximation takes the form:
\begin{gather}\label{N_cutting_number}
N \sim \frac{l q_v}{2\pi} =\frac{2p l}{\pi l_v}\sqrt{2\sum\limits_{m=0}^\infty \frac{1-A_q}{(2m+1)^2}} \ ,
\end{gather}
where we used the definition of $l_v$ given by \eqref{l_alpha} and $p=\sqrt{\ln(\tilde\kappa)/2\pi^3}$.

Performing the summation over $m$ for divergent terms we have the following expression for the equilibrium vortex density $n_{eq}$ and the vortex-dependent part of energy $E_{nV}$:
\begin{multline}\label{rho_min_simple}
n_{eq}(x)=\frac{16 M_0}{\Phi_0}\left[\left(1-\alpha\right)\sum\limits_{m=0}^{\infty}\frac{1-A_q}{2m+1}\sin(q x)-\right.\\\left.-
\frac{\pi^2 l_v^2}{l^2}\beta S_N\left(\frac{\pi x}{l}\right)\right] \ .
\end{multline}
\begin{multline}\label{E_nV_min_simple}
E_{nV} =\frac{E_M}{l^2}\sum\limits_{m=0}^\infty\left[\frac{1}{q_z^2}-\frac{q}{q_z^2 C_q}-\left(1 - \alpha\right)^2\frac{1-A_q}{q^2}\right]-\\ - E_M N\left(\frac{ l_v^2}{8 \lambda l}\right)^2 \ ,
\end{multline}
with
\begin{multline*}
S_N(z)=\sum_{m=0}^N (2m+1)\sin\left[(2m+1)z \right]=\\=
%\frac{\beta}{2}\left[\frac{\sin\left[(2N+3)z\right]}{\sin^2z}-\frac{(2N+3)\cos\left[(2N+2)z\right]}{\sin z}\right]
\frac{(2N+3)\sin\left[(2N+1)z\right]-(2N+1)\sin\left[(2N+3)z\right]}{4\sin^2z},
\end{multline*}
$\alpha=l_v^2/(8\lambda^2)=M_{th}^\lambda/M_0$%\frac{l_v^2}{8\lambda^2}
, and $\beta=\frac{\ln\tilde\kappa}{\ln(l/\xi)}$.

Comparing Eqs.~(\ref{E_MH}, \ref{E_nV_min_simple}) one can see that the first and the second terms in brackets of \eqref{E_nV_min_simple} compensate the corresponding terms in $E_{MH}$, therefore the volume energy in the vortex state takes the form:
\begin{multline}\label{E_nV+E_MH}
E_{vol}=E_M\left[\frac{\alpha(2-\alpha)}{8}-N\left(\frac{ l_v^2}{8 \lambda l}\right)^2 +\right. \\ \left.+
\sum\limits_{m=0}^\infty\frac{\left(1 - \alpha\right)^2 A_q}{q^2l^2}\right] \ ,
\end{multline}
with $A_q=q_z^2/\left[\lambda^{-2}+q q_z L_z(q_z+q\coth (q_z L_z))\right]$.

%%%%%%%%%%%%%%%%%%%%%%%%%%%
\begin{figure}[t]
\centerline{
\includegraphics[width=0.4\linewidth]{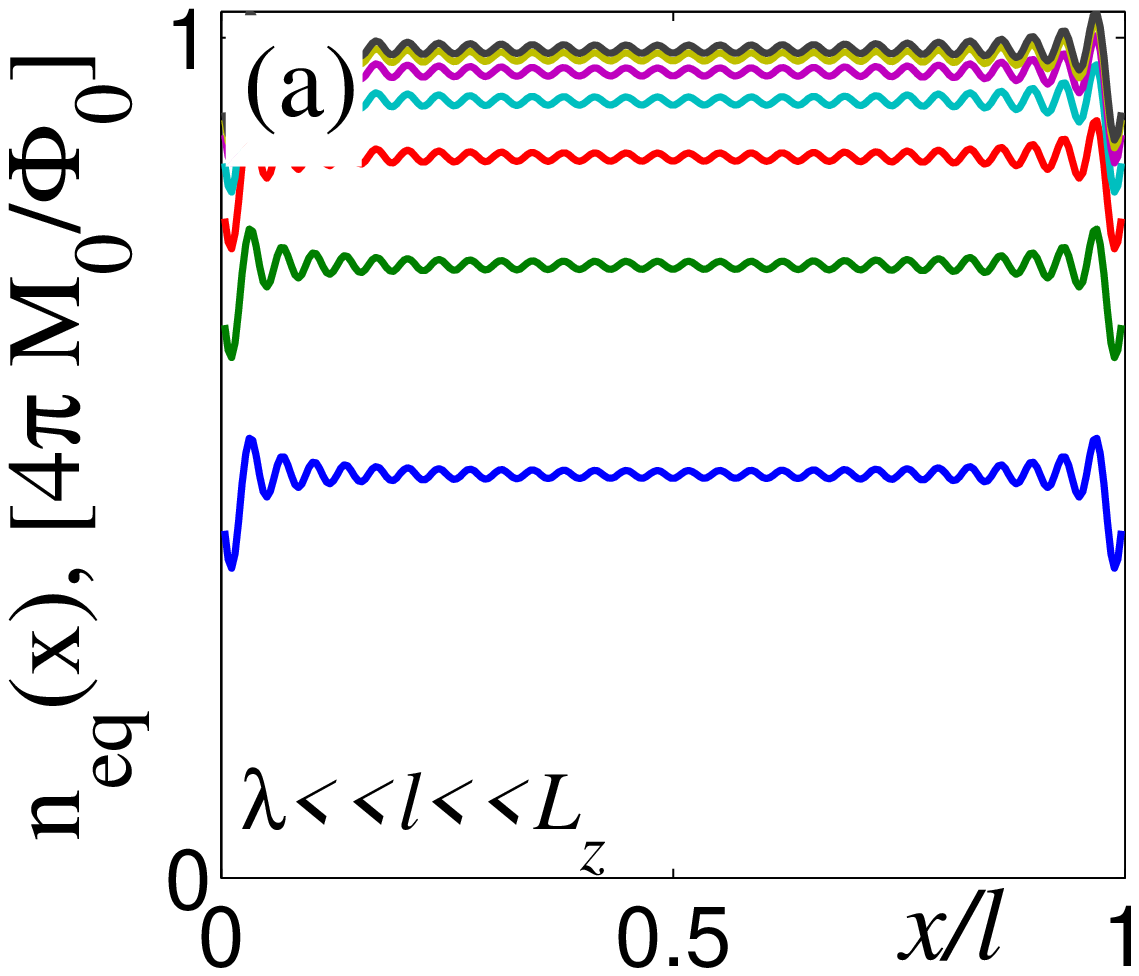}
\includegraphics[width=0.4\linewidth]{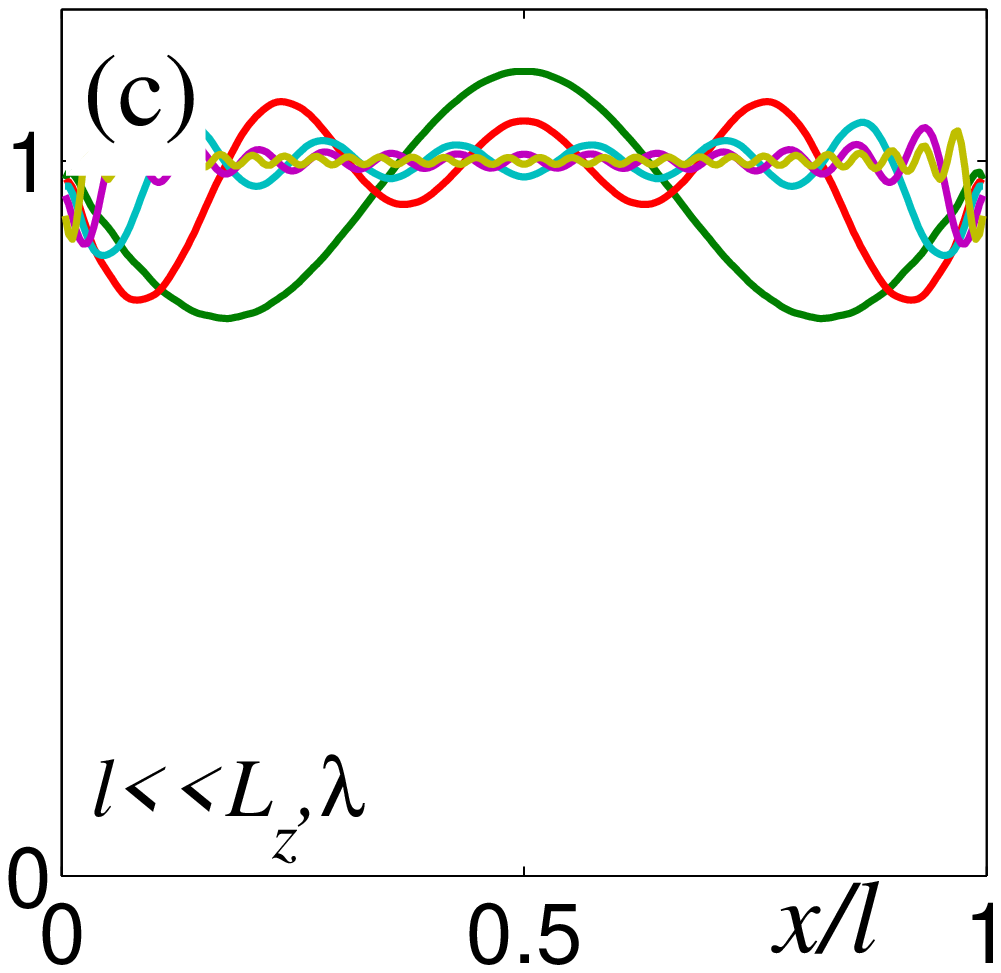}
}
\centerline{
\includegraphics[width=0.4\linewidth]{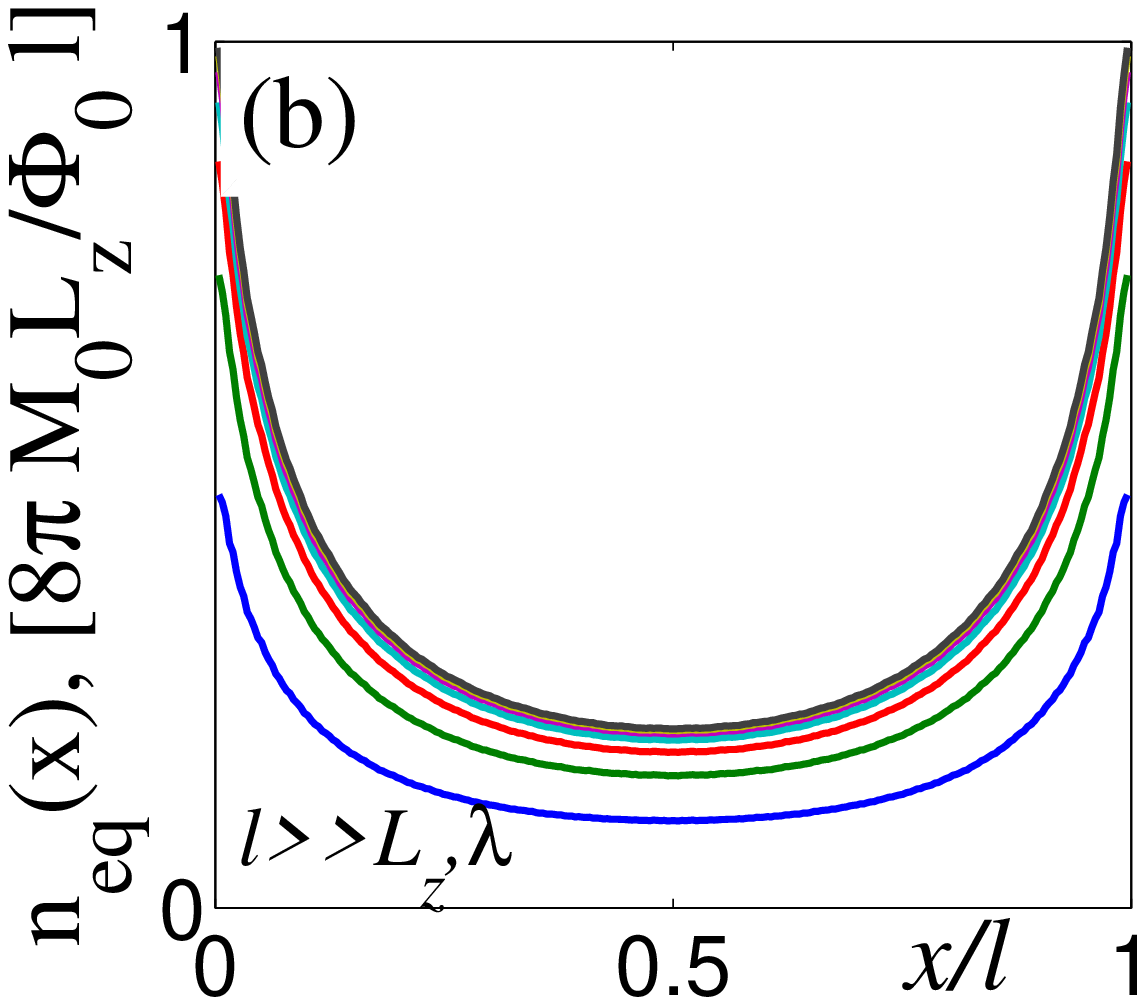}
\includegraphics[width=0.4\linewidth]{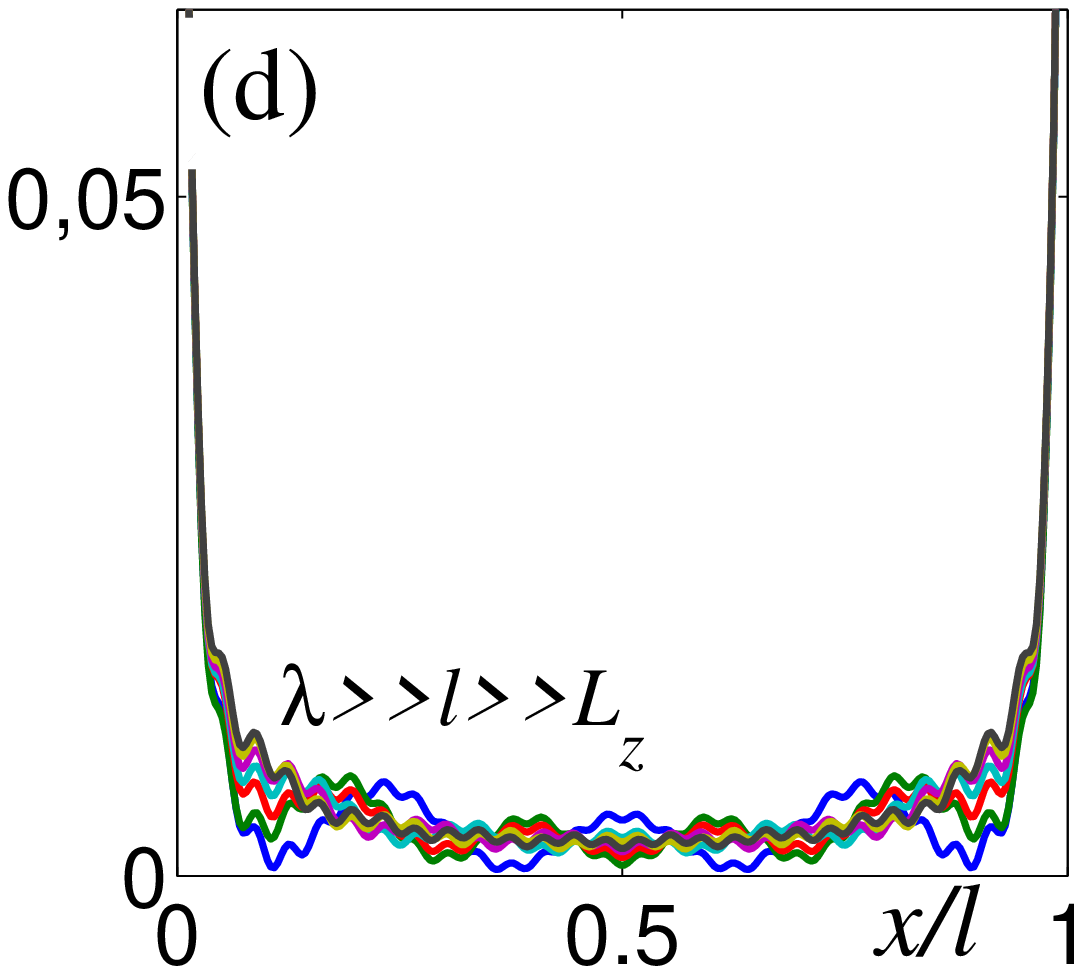}
}
\caption{The space profiles of the vortex density distribution function for the different parameter values: (a) $l_v=10^{-4}L_z$, $l=0.01 L_z$, (b) $l_v=0.01 L_z$, $l=10 L_z$; the plots at panels (a, b) from bottom to top correspond to the increasing ratio $\lambda/l_v$ from the threshold value $2^{-3/2}$ to the tenth higher value; (c) $l_v=10^{-4}L_z$, the plots from bottom to top correspond to the increasing ratio $l/l_v$ from the value $1.6$ to the value $10$; (d) $l_v=3 L_z$, the plots from bottom to top correspond to the increasing ratio $l/l_v$ from the value $\sim 80$, where $n_{eq}(x)$ becomes positive, to the tenth higher value. Note that plots at panels (c, d) remain intact for any $\lambda\gg l$.
%Panels (a) and (c) correspond to Eq.~\eqref{rho_L<<1}, while  (b) and (d)~---~to Eq.~\eqref{rho_L>>1}}
}
\label{Fig:SCferro_rho}
\end{figure}
%%%%%%%%%%%%%%%%%%%%%%%%%%%

The expressions for the vortex density \eqref{rho_min_simple} and for the volume energy per unit area \eqref{E_nV+E_MH} in the vortex state at equilibrium are valid in the following assumptions, mentioned above:
(i) the calculated vortex density is assumed to be positive $n_{eq}(x)>0$ in the range $0<x<l$ for applicability of the expression for the (anti)vortex number \eqref{N_V};
(ii) the continuous approximation of $n(x)$ restricts our consideration of the vortex state to the range of parameters far from the vortex penetration threshold, i.e. $M_0\gg M_{th}(l,\lambda,L_z)$ and/or $N\gg 1$.

The latter condition results in the fact that our model gives the best results describing the vortex penetration in the case of strong shrinking of domains in the Meissner state\cite{Buzd_Dao} ($\tilde w\ll\lambda\ll l\ll L_z$), due to the equilibrium domain size $l$ in the Meissner state $l_S$ is small compared with the one $l_S^v$ in the vortex state $l_S\ll l_S^v$ and we can use the expression \eqref{E_nV+E_MH} far from the threshold $l_v\ll l_v^{th}\simeq l_S^v$ \eqref{alph_th_L<<Lmd} for $l\ll L_z$.
Note that the condition (i) can be weakened due to the fact that using the expression \eqref{N_V} for the number of (anti)vortices in the sample we can only underestimate the volume energy in the vortex state by Eq.~\eqref{E_nV+E_MH}.

In the rest of the section we will present the vortex density distribution function $n_{eq}(x)$ at equilibrium state.
Keeping for simplicity the main terms in \eqref{rho_min_simple} we will focus on the thick $L_z\gg l$ and the thin $L_z\ll l$ FS films:

(i) For the thick samples where vortices interact as the ones in the bulk SC we can neglect the terms with $A_q$ in \eqref{N_cutting_number} and \eqref{rho_min_simple} comparing with unity, due to $q L_z\gg1$ and therefore $A_q<1/(q L_z)\ll 1$, and obtain
\begin{gather}\label{rho_L<<1}
n_{eq}(x)\approx \frac{4\pi}{\Phi_0}\left[M_0-M_{th}^{\lambda}-\frac{\pi}{2}M_{th*}S_N\left(\frac{\pi x}{l}\right)\right]\ ,
\end{gather}
with the maximum value of $S_N(z)$ of order of $S_N^{max}\sim N^2$ at $z\sim 1/\pi N$, $M_{th*}$, $M_{th}^\lambda$ are given by \eqref{M_th_L_z,lmd_gg_l} and \eqref{M_th_lmd_ll_l}, respectively, and
$N\approx p l/l_v$ with $p$ given after Eq.~\eqref{N_cutting_number}. %=\sqrt{\ln(\tilde\kappa)/2\pi^3}$.
The typical vortex density profiles demonstrated on Figs.~\ref{Fig:SCferro_rho}(a,~c)
 for different ratios $l/\lambda$ and $l/l_v$ are nearly constant far from the threshold $M_0=M_{th}$.

For rather low temperatures $l\gg\lambda(T)$, when the domain size $l$ is large compared with penetration depth $\lambda$
the vortices penetrate the sample with almost homogeneous density straight above the penetration threshold. Their density changes only in amplitude with
increasing $\lambda/l_v$ ratio (see Fig.~\ref{Fig:SCferro_rho}(a)).
The deviations from this average vortex density originate from the last small term in \eqref{rho_L<<1}.

Unlike this in the vicinity of the superconducting phase transition $T\lesssim T_c$, when $l\ll\lambda(T)$, the latter term  in \eqref{rho_L<<1} crucially changes the space profile of the vortex density (see Fig.~\ref{Fig:SCferro_rho}(c)). Straight above the threshold $M_0\gtrsim M_{th}^\lambda$ the only one (anti)vortex enters each domain, while far from the threshold the above-mentioned constant vortex density $n_{eq}(x)=4\pi M_0/\Phi_0$ remains.

(ii) In rather thin FS films the vortices appear as the Pearl-like structures with long-range repulsive interaction, which leads to the vortex concentration near the domain walls.
%vortices repel more effectively, due stray field induced by their the Pearl-like structure.
%In this case one can rewrite $A_q$ as follows
%\begin{gather}
%A_q = \frac{1}{1+q L_z+[q_z L_z\coth(q_z L_z)-1]q^2/q_z^2} \ ,
%\end{gather}
%where $[q_z L_z\coth(q_z L_z)-1]/(q_z L_z)^2<1/3$.
Indeed, using the fact that in this case the sum in \eqref{rho_min_simple} converges at $q\lesssim L_z^{-1}$ one can approximate the main term as $(1-A_q)/(q L_z)\approx 1/(1+q L_z)\approx 1$ for $A_q\approx1/(1+q L_z)$. As a result we have
\begin{gather}\label{rho_L>>1}
n_{eq}(x)\approx \frac{8\pi L_z}{\Phi_0 l}\left[\frac{M_0-M_{th}^\lambda}{\sin(\pi x/l)}-\frac{8 G}{\pi}M_{th}^* S_N\left(\frac{\pi x}{l}\right)\right]\
\end{gather}
and $N\approx 2 p \sqrt{l L_z\ln(l/L_z)}/(\pi l_v)$. Here $M_{th}^*$, $M_{th}^\lambda$ are given by \eqref{M_th_L_z_ll_l_ll_lmd} and \eqref{M_th_lmd_ll_l}, respectively.
One can see that the typical vortex density profiles shown at Figs.~\ref{Fig:SCferro_rho}(b, d) for different ratios $l/\lambda$ and $l/l_v$ have the maxima near the domain walls.
For rather low temperatures the spatial distribution of $n_{eq}(x)$ is similar to the case, considered by Erdin and coauthors in superconductor-ferromagnet bilayer.\cite{Pokrovsky} Straight after the SC phase transition the vortex density is inversely proportional to the $\sin$-function in each domain and increasing of the ratio $\lambda/l_v$ far from the threshold value $\left(\lambda/l_v\right)_{th}=2^{-3/2}$ only scales the vortex density.

Unfortunately, for rather large temperatures, $l\ll\lambda(T)$, our model fails to get the vortex distribution function correctly for the parameters not far from the threshold, due to the negative values of $n_{eq}(x)$ in this case, which contradicts to the assumption that the only vortices are in the domain $0<x<l$.
The failure of the model seems to be originated from rather inhomogeneous vortex density $n_{eq}(x)$ with rare vortices in the middle of the domains
and from the rude approximation of $n_{eq}(x)$ by a finite amount of Fourier harmonics.
In the Fig.~~\ref{Fig:SCferro_rho}(d) one can see that the vortices for the parameters far from the threshold are mainly situated near the domain walls as in the low temperature case.
As we will see in the next section such a problem with calculation of the equilibrium vortex distribution leads to the failure of consideration of the Meissner/vortex phase transitions in this limit.

Summarizing, we point out that in all the cases the equilibrium vortex distribution far from the penetration threshold is determined by the intervortex interaction depending on the FS film thickness, while the transformation of $n_{eq}(x)$ not far from the threshold is mainly governed by the stray field screening parameters.

%%%%%%%%%%%%%%%%%%%%%%%%%%%%%%%%%%%%%%%%%%%%%%%%%%%%%%%%%%%%%%%%%%%%%%%%%%%%%%%%%%%%
\section{Energetically favorable domain configurations. First order phase transitions.}\label{sec5_eq_energy}
In this section we analyze the transformation of the ground state of the FS film shown at the energy diagram in Fig.~\ref{Fig:SCferro_stripe-structure}(d) with decreasing temperature $T$. Without loss of generality we put the minimum value of the London penetration depth $\lambda(0)$ at zero temperature to be small compared with the film thickness $L_z$, the domain size $l$, and the effective domain wall width $\tilde w$.

During this analysis we assume that both the pinning potentials for vortices and domain walls and the Bean-Livingston barrier are negligibly small and consider the system phase transitions in terms of the diagram in Fig.~\ref{Fig:SCferro_stripe-structure}(d). In other words we will calculate the minimum values of the FS film total energy in the Meissner and the vortex states using Eqs.~\eqref{E_MH} and \eqref{E_nV+E_MH}, respectively, and will compare which of the states realizes the global minimum of energy. Using the vortex penetration threshold value derived in Sec.~\ref{sec3_BL_barrier} we can check if the Meissner state is locally stable for the small fluctuations of the vortex number. The local stability of the vortex state can be verified within our model only for rather dense vortex lattices $N_v/(l L_y)\gg 1$.

The equilibrium domain size $l_S$ [$l_S^v$] in the Meissner [vortex] state of the sample is obtained by minimization of the total energy $E(l)=E_{DW}(l)+E_{MH}(l)$ [$E(l)=E_{DW}(l)+E_{MH}(l)+E_{nV}(l)$].
For the temperatures $T>T_c$ larger than the superconducting critical temperature $T_c$ the FS film remains in the normal state without emerging of the superfluid screening currents.
The volume energy in the this case can be obtained from Eq.~\eqref{E_MH} in the limit $\lambda\to\infty$ or from Eq.~\eqref{E_nV+E_MH} in the limit $l_v\to0$ and $\lambda\to\infty$. Both these limits lead to the same equilibrium DS \cite{Kittel,LL,Buzd_Dao} with the following asymptotics:
\begin{gather}\label{L_N_<<1}
l_N=\sqrt{\frac{\pi^2 \tilde w L_z}{14\zeta(3)}} \ll L_z
\end{gather}
for $\tilde w\ll L_z$ and
\begin{gather}\label{L_N_>>1}
l_N=\pi L_z\exp[\tilde w/16 L_z-1/2] \gg L_z
\end{gather}
for $\tilde w\gg L_z$.

Further we will use the results of the calculation of the domain size minimizing the total energy in the Meissner state, which has been previously done in the papers [\onlinecite{Buzd_Faure,Buzd_Dao}].
To calculate the equilibrium DS in vortex state we discuss below the limiting cases in details. Similarly to Ref.~\onlinecite{Buzd_Dao} we consider normalized domain size $L=l/L_z$ and momenta $Q=q L_z$, $Q_z=q_z L_z$ and obtain for the limiting cases:

(i) For rather thick FS films $L_z\gg l$ where domain size $l$ is essentially smaller than the film thickness $L_z$ we obtain
%\begin{multline}\label{A_Q_L<<1}
$A_q \approx {1}/{2 b Q} $, %\ ,
%\end{multline}
with $b=(1+Q/Q_z)/2$. In this case all the terms in Eq.~\eqref{E_nV+E_MH} converge at $m\sim 1$ with integer $m$ defined by $Q=\pi (2m+1)/L$, i. e. one can write the following expression for volume energy in the vortex state for $l\ll L_z$ %($l_v^{th}\approx l$):
\begin{multline}\label{E_nV+E_MH_L<<1}
E_{vol} \approx \frac{E_M}{8}\left[%1-\left(1-\alpha\right)^2
\alpha(2-\alpha)
-\frac{p l_v^3}{8 \lambda^2 L_z}\frac{1}{L}+\right.\\ \left.+\left(1-\alpha\right)^2\frac{7\zeta(3) L}{2 b \pi^3}\right] \ ,
\end{multline}
where
we approximate the parameter $b$ by its value at $m=1$: $b\approx\left[1+\pi\lambda/\sqrt{\pi^2\lambda^2+l^2}\right]/2$ and
$N\approx p l/l_v$ with expression for $p$ given after Eq.~\eqref{N_cutting_number}.%=\sqrt{\ln(\tilde\kappa)/2\pi^3}$

(ii) In the opposite case of the thin samples $L_z\ll l$ one can use the expression for the last term in the first line of Eq.~\eqref{E_nV_min_simple}
$$
\frac{1-A_Q}{Q^2}= \frac{1}{Q+1}\left(\frac{1}{Q}+\frac{g[Q_z]}{1+Q+g[Q_z] Q^2}\right)%\frac{1+f Q}{Q(1+Q+f Q^2)}
\ ,$$
with $g[x] = (x \coth x-1)/x^2$. The function $g[Q_z]$ is proportional to $\lambda/L_z$ for $\lambda\ll L_z$ and is close to $1/3$ for $\lambda\gg L_z$.

For rather low temperatures, when $\lambda(T)\ll L_z$, one can approximate the last term in the brackets as $\lambda/ \left[L_z(1+Q)\right]$.
Therefore for $\lambda\ll L_z\ll l$ the volume part of energy can be written as follows
\begin{multline}\label{E_nV+E_MH_Lmd<<1<<L}
E_{vol} \approx E_M\left[\frac{1}{8}-\frac{(1-\alpha)^2}{2\pi L}\left[{\ln(L/\pi)+\gamma}\right]\right]-\\-
E_M\frac{\lambda}{L_z}\left[\frac{(1-\alpha)^2}{2\pi L}+\frac{p \alpha^{3/2} }{\sqrt{2}\pi}\frac{\sqrt{\ln L}}{L^{3/2}}\right]
\end{multline}
with $N\approx 2 p \sqrt{l L_z\ln(L)}/(\pi l_v)$ and Euler constant $\gamma\approx 0.57722$.

In the vicinity of the superconducting phase transition $T\lesssim T_c$, when $L_z\ll\lambda(T)$ (cf. the previous section),
one can expand function $g$ over small parameter $L_z/\lambda$ as follows $g[Q_z]\approx g[Q]+ g'[Q] \frac{L_z^2}{2\lambda^2 Q}$ using the fact that $g'[x]/(2 x g[x])<0.2$ for any $x$, therefore
\begin{multline*}%\label{A_Q_L<<Lmd}
A_q \approx \frac{1-e^{-2Q}}{2Q}+\\+\frac{1}{2 \lambda^2 q^2}\left[e^{-Q}+\frac{1-e^{-4Q}}{4Q}-\frac{(1-e^{-2Q})^2}{2 Q^2}\right] \ ,
\end{multline*}
and one can write the following expression for volume energy in the vortex state for $L_z\ll l,\lambda$ %($l_v^{th}\approx \sqrt{8 G l L_z/\pi^2}$)
\begin{multline}\label{E_nV+E_MH_1<<L,Lmd}
E_{vol} \approx E_M\left[\frac{1}{8}-\left(1-\alpha\right)^2 \frac{\ln(L/\pi)+3/2}{2\pi L}\right]+\\  +
E_M\frac{L_z^2}{\lambda^2}\left[(1-\alpha)^2\frac{c}{\pi L}-\frac{p l_v^3 }{32\pi L_z^3}\frac{\sqrt{\ln L}}{L^{3/2}}\right] \ ,
\end{multline}
with $c=(17/4-6\ln2)/15\approx 0.006$.

Comparing the minimum values $\epsilon_S$ and $\epsilon_v$ of the total energy normalized to $E_M$ in the Meissner and the vortex states, respectively, using Eq.~\eqref{E_MH} and Eqs.~(\ref{E_nV+E_MH_L<<1}, \ref{E_nV+E_MH_Lmd<<1<<L}, \ref{E_nV+E_MH_1<<L,Lmd}) one can calculate the domain size value which realize the global minimum of the total energy $(E_{DW}+E_{vol})/E_M$ for different amplitudes of magnetization $M_0$.
As was mentioned in the Sec.~\ref{sec1_intro} in the typical situation for rather large $\lambda(T)$ the vortex state is stable, while the Meissner state is fully unstable for any DS value, and the DS at equilibrium $l_S^v$ is almost equal the normal DS $l_S^v\approx l_N$.
As it will be discussed below with decreasing temperature at the certain critical London penetration depth $\lambda_S$ the Meissner state becomes locally stable, i.e. the vortex penetration threshold $M_{th}(\lambda_S,l_S(\lambda_S))$ at the DS $l=l_S$ realizing the minimum of energy in the Meissner state becomes higher than the magnetization $M_0$. But the vortex state still realizes the global minimum of the energy. For $\lambda<\lambda_m<\lambda_S$ smaller than another critical value $\lambda_m$ [when $\epsilon_S(l_S(\lambda_m))=\epsilon_V(l_S^v(\lambda_m))$] the total energy minimum of the FS film realizes in the Meissner state $E/E_M=\epsilon_S(l_S)$ with the domain size $l=l_S$.
Note that in general case $\lambda_m\ne\lambda_S$. As we will see below the described scenario corresponds to the type I phase transition between the vortex and the Meissner states in the case $\lambda_m<\lambda_S$ even for neglecting the effects of the Bean-Livingston barrier.

Note that we can't use Eqs.~(\ref{E_nV+E_MH_L<<1}, \ref{E_nV+E_MH_Lmd<<1<<L}, \ref{E_nV+E_MH_1<<L,Lmd}) as exact energy profiles near the vortex penetration threshold, due to breakdown of the approximation \eqref{N_V}.
%using the expression \eqref{N_V} for the number of (anti)vortices in the domain.
However, we can underestimate the volume energy in the vortex state by Eq.~\eqref{E_nV+E_MH}.
As a result comparing the minimum values $\epsilon_S(l_S)$ and $\epsilon_v(l_S^v)$ of free energies in the Meissner and the vortex states in this case
in addition to calculation of the Meissner state instability threshold $\lambda_S$ we can make lower estimate of $\lambda_m$ by $\epsilon_S(l_S)<\epsilon_v(l_S^v)$.

For clarity of presentation further we consider the regimes of large ($\tilde w\gg L_z$) and small ($\tilde w\ll L_z$) values of the effective domain wall width comparing with the FS film thickness $L_z$ separately. As we know from Eqs.~(\ref{L_N_<<1}, \ref{L_N_>>1}) these cases correspond to the large ($l_N\gg L_z$) and small ($l_N\ll L_z$) values of the DS in the normal state, respectively.

\subsection{Thick films with $L_z\gg l_N$.}
When the effective domain wall size $\tilde w$ is small compared with the FS film width $L_z$ the Meissner screening currents can change the equilibrium DS $l$ crucially for certain penetration depth values.
For weak magnetization amplitudes $M_0<M_{th*}(l_N)$ with $M_{th*}(l)$ given by \eqref{M_th_L_z,lmd_gg_l} the vortex penetration threshold can't be reached for any $\lambda$ values.
In other words for these magnetization values the Meissner state realizes the equilibrium state and scenario of domain shrinkage discussed in [\onlinecite{Buzd_Dao}] takes place.

Using the results of [\onlinecite{Buzd_Dao}] one can write down the expression for the Meissner state free energy
in the vicinity of the superconducting phase transition $\lambda\gg l$
\begin{gather}\label{E_MH_L<<1,Lmd}
\epsilon=\frac{E_{DW}+E_{MH}}{E_M}=\frac{7\zeta(3)}{16\pi^3}\left[L+\frac{L_N^2}{L}\right]+\frac{L^2}{96\Lambda^2} \ ,
\end{gather}
with $\Lambda=\lambda/L_z$.

The the minimum value of Eq.~\eqref{E_MH_L<<1,Lmd} for $\lambda\gg (l_N L_z)^{1/2}$
\begin{gather}\label{E_S^*_L<<1}
\epsilon_S(l_S^*)\approx\frac{7\zeta(3)L_N}{8\pi^3}+\frac{L_N^2}{96\Lambda^2} \ .
\end{gather}
realizes at the equilibrium DS
\begin{gather}\label{L_S^*}
l_S^*=l_N\sqrt{1-\frac{\pi^3 l_N L_z}{21\zeta(3)\lambda^2}}\simeq l_N \ ,
\end{gather}
while in the opposite limit of $\tilde w\ll\lambda\ll (l_N L_z)^{1/2}$ the minimization of free energy \eqref{E_MH_L<<1,Lmd} leads to
\begin{gather}\label{E_S*_L<<1}
\epsilon_S(l_{S*})\approx\frac{L_{S*}^2}{32\Lambda^2}+\frac{7\zeta(3)L_{S*}}{8\pi^3}\approx\frac{L_{S*}^2}{32\Lambda^2}
\end{gather}
reaching for rather small domain width
\begin{gather}\label{L_S*}
l_{S*}=\left(\frac{21\zeta(3)\lambda^2 l_N^2}{\pi^3 L_z}\right)^{1/3}\ll l_N,\lambda \ .
\end{gather}
Further decreasing of the penetration depth $\lambda$ results in the decreasing of domain size to the minimum value $l_{min}\approx 0.59 (\tilde w/\pi)$ at $\lambda_{min}\approx k\cdot\lambda_c$ with $\lambda_c=\tilde w/(8\pi)$ and $k\sim 1.3-1.6$. In the range $\lambda_c\lesssim\lambda<\lambda_{min}$ the DS increases with decreasing $\lambda$ as follows:\cite{Buzd_Dao}
\begin{gather}\label{L_S_1,Lmd<<L}
l_S =\frac{\pi\lambda^2}{L_z^{1/2}\left[\pi \lambda - 2I(\lambda) - \tilde w/8\right]^{1/2}} \ ,
\end{gather}
with
\begin{gather}\label{I(Lmd)}
I(\lambda)=\int\limits_0^\infty \frac{L_z q dq}{q_z^3(q_z+q\coth q_z)} \ ,
\end{gather}
which takes the form $I(\lambda)\approx \lambda^2(1-\ln2)/L_z\ll\lambda$ at $\tilde w\ll L_z$.
For $\lambda\lesssim\lambda_c$ the FS film goes to the monodomain state ($l\to L_x\gg L_z$ in considering case) \cite{Sonin,Buzd_Faure,Buzd_Dao}.

If the magnetization value is larger than the critical one $M_0>M_{th*}(l_N)$ vortices penetrate the sample and realize the global minimum of the energy for $\lambda>\lambda_m$. Moreover the vortex state is the only stable one for $\lambda>\lambda_S$.
One can minimize the total energy $\epsilon=(E_{DW}+E_{vol})/E_M$ in the vortex state (when it is stable) with $E_{vol}$ given by \eqref{E_nV+E_MH_L<<1} and obtain the minimal value
\begin{multline}\label{E_v_L<<1}
\epsilon_v(l_S^v)= \frac{7\zeta(3) L_N }{8\pi^3}\sqrt{\frac{(1-\alpha)^2}{b}\left(1-\frac{\pi^3 l_N L_z}{21 \zeta(3)\lambda^2}  r_N \right)}+\\+
\frac{\alpha(2-\alpha)}{8}\ ,
\end{multline}
realized at the domain size
\begin{gather}\label{L_S^v_L<<1}
l_S^v=l_N\sqrt{\frac{b}{(1-\alpha)^2}\left(1-\frac{\pi^3 l_N L_z}{21 \zeta(3)\lambda^2}  r_N \right)} \ ,
\end{gather}
with $\alpha=l_v^2/8\lambda^2$, $b\approx\left[1+\pi\lambda/\sqrt{\pi^2\lambda^2+l_N^2}\right]/2$, and small parameter $r_N={3 p l_v^3}/(4 l_N^3)\ll 1$.
%Here we introduce the notations for the phase transition penetration depth $\Lambda_{m}$, when the minimum energy values in the Meissner $E_S(L_S)$ and in the vortex $E_V(L_S^v)$ state are equal $E_S(L_S(\Lambda_m))=E_V(L_S^v(\Lambda_m))$, and the threshold of the Meissner state stability $\Lambda_S$, when the vortices can penetrate the sample at the DS $L=L_S$ which realizes the minimum of energy in the Meissner state $M_0=M_{th}(\Lambda_S,L_S(\Lambda_S))$. Note that in general case $\Lambda_m\ne\Lambda_S$. As we will see below the transformation between the Meissner and the vortex states is the type I phase transition in the case $\Lambda_m<\Lambda_S$.
Further we consider some typical scenarios of domain size variation with decreasing $\lambda(T)$ for different magnetization values.

(i) In the vicinity of the critical magnetization value $M_0/M_{th*}(l_N)-1\ll1$
the vortex penetration threshold is achieved in vicinity of the superconducting phase transition, with the relatively small domain size shrinkage $\lambda\gg (l_N L_z)^{1/2}$, see \eqref{L_S^*}. In this case the vortex state realizes at least at $\lambda>\lambda_S$, when the Meissner state becomes unstable, with
\begin{gather}\label{Lmd_S^*}
\lambda_S = \left[\frac{\pi^3 l_N L_z}{21\zeta(3)(1-l_v^2/l_N^2)}\right]^{1/2}
\end{gather}
derived from \eqref{M_th_L_z,lmd_gg_l} $l_S^*(\lambda_S)=l_v$.

Using Eqs.~\eqref{L_S^*} and \eqref{L_S^v_L<<1} one can easily obtain that both DS in the vortex
\begin{gather*}%\label{L_S^v*_L<<1,Lmd}
l_S^v\approx l_N\sqrt{1-\frac{\pi^3 l_N L_z}{21\zeta(3)\lambda^2}r_N} \sim l_N
\end{gather*}
and the Meissner states
$l_S^*\sim l_N$ are very close to its normal state value. Therefore in this case the vortex state becomes globally stable near the penetration threshold and we can't compare the energy profiles $\epsilon_S(L)$ and $\epsilon_V(L)$ in this case to say explicitly if the phase transition between the vortex and the Meissner states is the type I or type II transition. The only information about the vortex state we can get is that for $\lambda\gg \lambda_S$ the vortex state is the only stable one and the minimum value of the total energy gives
\begin{gather*}%\label{E_v^*_L<<1}
\epsilon_V(l_S^v)\approx
\frac{7\zeta(3)L_N}{8\pi^3}+\frac{L_N^2}{96\Lambda^2}\frac{3 M_{th*}}{M_0}\left[1-\frac{p}{2}\sqrt{\frac{M_{th*}}{M_0}}\right] \ ,
\end{gather*}
derived from \eqref{E_v_L<<1}.

For $\lambda<\lambda_S$ the Meissner state is surely stable if the lower estimation of the minimum free energy value in the vortex state given by latter expression is larger than the one in the Meissner state \eqref{E_S*_L<<1}. This inequality can be reached for some $l<l_N$ at $M_0<3M_{th*}(l)$ for $p\ll 1$ and $M_0<M_{th*}(l)$ for $p\lesssim 4/3$.
For lower penetration depths the Meissner state realizes the global energy minimum.
\begin{figure}[t]
\centerline{
\includegraphics[width=0.9\linewidth]{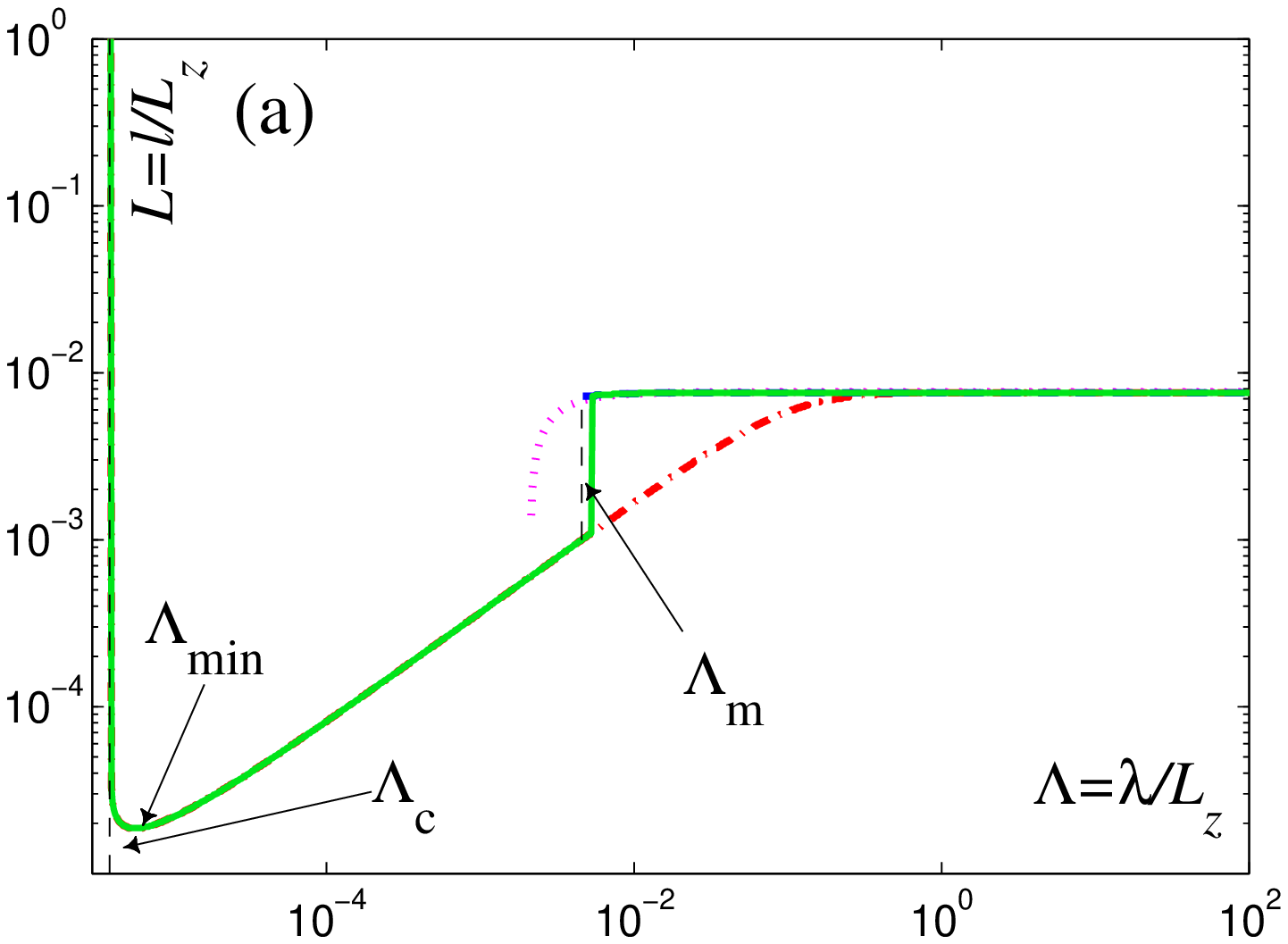}
}
\centerline{
\includegraphics[width=0.88\linewidth]{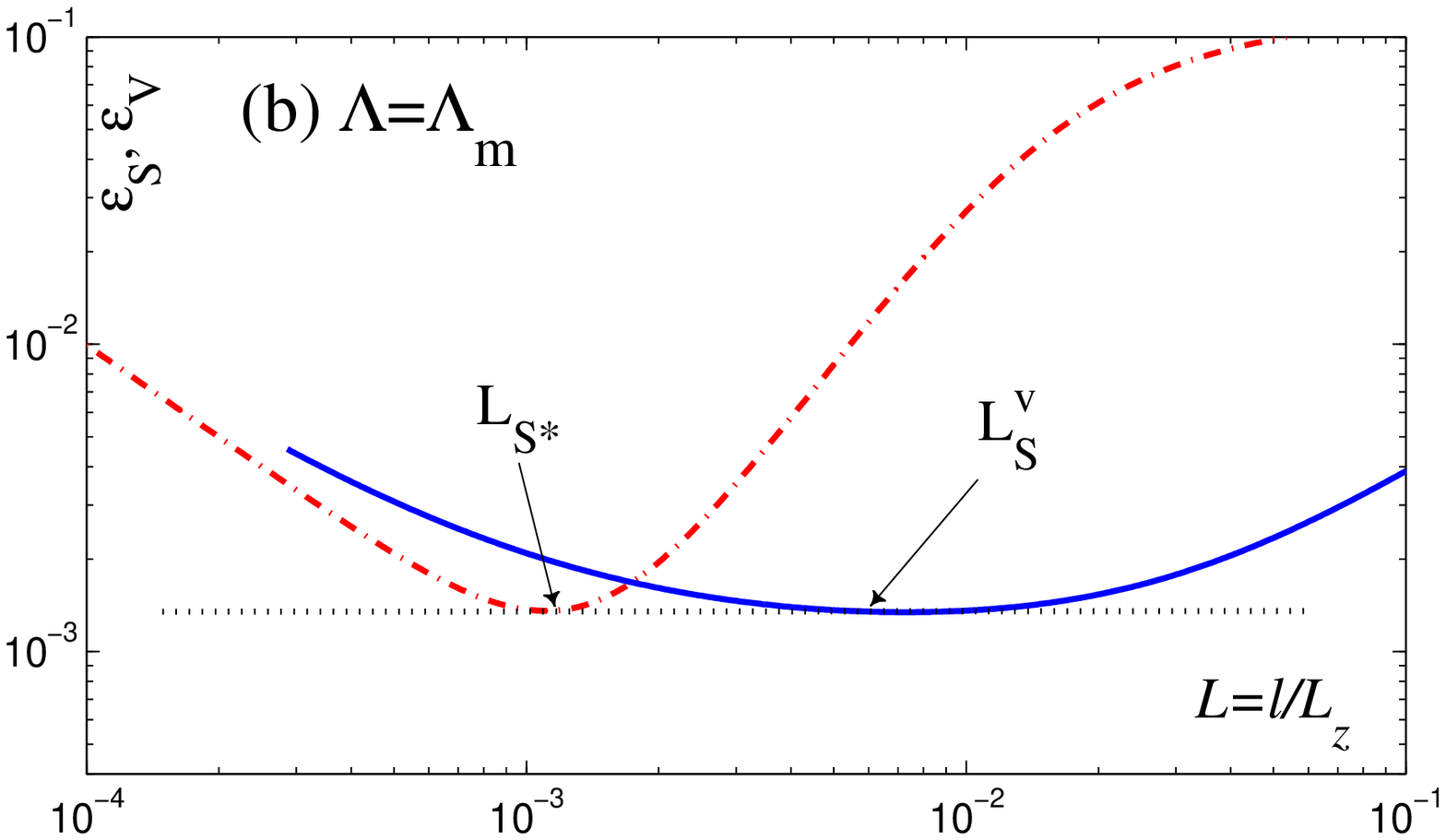}
}
\caption{(a) Domain size $l$ normalized to the film thickness $L_z$ at equilibrium state vs the penetration depth $\Lambda=\lambda/L_z$. The solid green line corresponds to the DS which realizes the global minimum of the total energy, while the dashed blue (dash-dotted red) line corresponds to the numerically calculated DS in the vortex (Meissner) state. The dotted pink line is the analytical DS in the vortex state given by \eqref{L_S^v_L<<1}.
The London penetration depths, where the state which realized total energy minimum changes ($\Lambda_m$), where the minimum of DS reached ($\Lambda_{min}$), and where the DS is diverged ($\Lambda_c$), are shown by arrows; (b) Comparative plot of the total energies in the vortex (solid blue line) and the Meissner (dash-dotted red line) states vs the domain size $l/L_z$ for $\lambda=\lambda_m$. For both plots parameters are following $\tilde w/L_z=10^{-4}$ and $l_v/L_z=10^{-3}$.}
\label{Fig:SCferro_Lmin_and_E_SV_W<<1}
\end{figure}

(ii)
In the most interesting case of $M_{th*}(l_N)\ll M_0\ll M_c$, where $M_c=4\Phi_0\ln\tilde\kappa/\tilde w^2$, the vortices realize the only minimum of the total energy for $\lambda>\lambda_S$ with
\begin{gather}\label{Lmd_S*}
\lambda_S = \left[\frac{\pi^3 l_v^3 L_z}{21\zeta(3)l_N^2}\right]^{1/2} \ .
\end{gather}
Note that $l\ll\lambda_S\ll\sqrt{l_N L_z}$, because for these $\lambda$ the vortices start penetrate to the sample even with the dense domain structure $l=l_{S*}$ \eqref{L_S*}.
Therefore for $\lambda>\lambda_S$ the equilibrium domain size $l=l_S^v$ can be rewritten from \eqref{L_S^v_L<<1} as follows ($b\approx 1$, $\alpha\ll 1$):
\begin{gather}\label{L_S*^v_L<<1,Lmd}
l_S^v=l_N\sqrt{1-\frac{3 p}{4}r_\lambda^3} \ ,
\end{gather}
with $r_\lambda=l_v/l_{S*} = (\lambda_S/\lambda)^{2/3}$ and the minimum total energy \eqref{E_v_L<<1}
\begin{gather}\label{E_v*_L<<1}
\epsilon_V(l_S^v)\approx\frac{7\zeta(3)L_N}{8\pi^3}\sqrt{1-\frac{3 p}{4}r_\lambda^3}+\frac{L_{S*}^2}{32\Lambda^2}r_\lambda^2\approx\frac{L_{S*}^2}{32\Lambda^2}r_\lambda^2 \ ,
\end{gather}
where the first term is small compared with the second one and can be neglected.
Note that our considerations works well only far from the threshold $l_S^v\gg\l_v$, i.e.,
$$\frac{l_{S*}^2}{l_N^2}r_\lambda^2\ll 1-\frac{3 p}{4}r_\lambda^3 \ ,$$
at least it leads to $r_\lambda^3<4/(3 p)$.
However, in the case of $l_S^v\gg l_{S*}$ we can describe the phase transition between the Meissner and the vortex states explicitly.
\begin{figure}[t]
\centerline{
\includegraphics[width=0.5\linewidth]{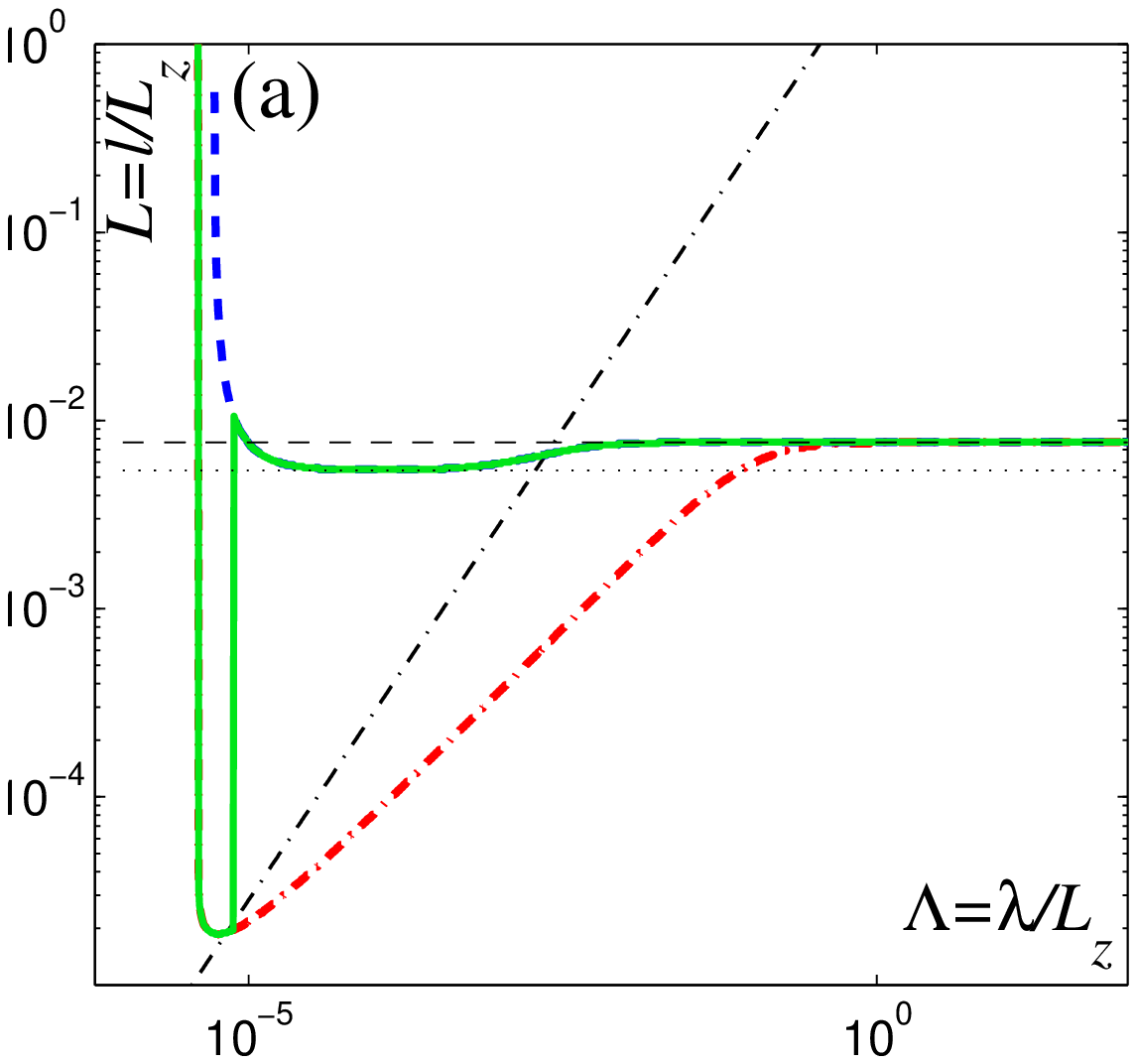}
%}
%\centerline{
\includegraphics[width=0.5\linewidth]{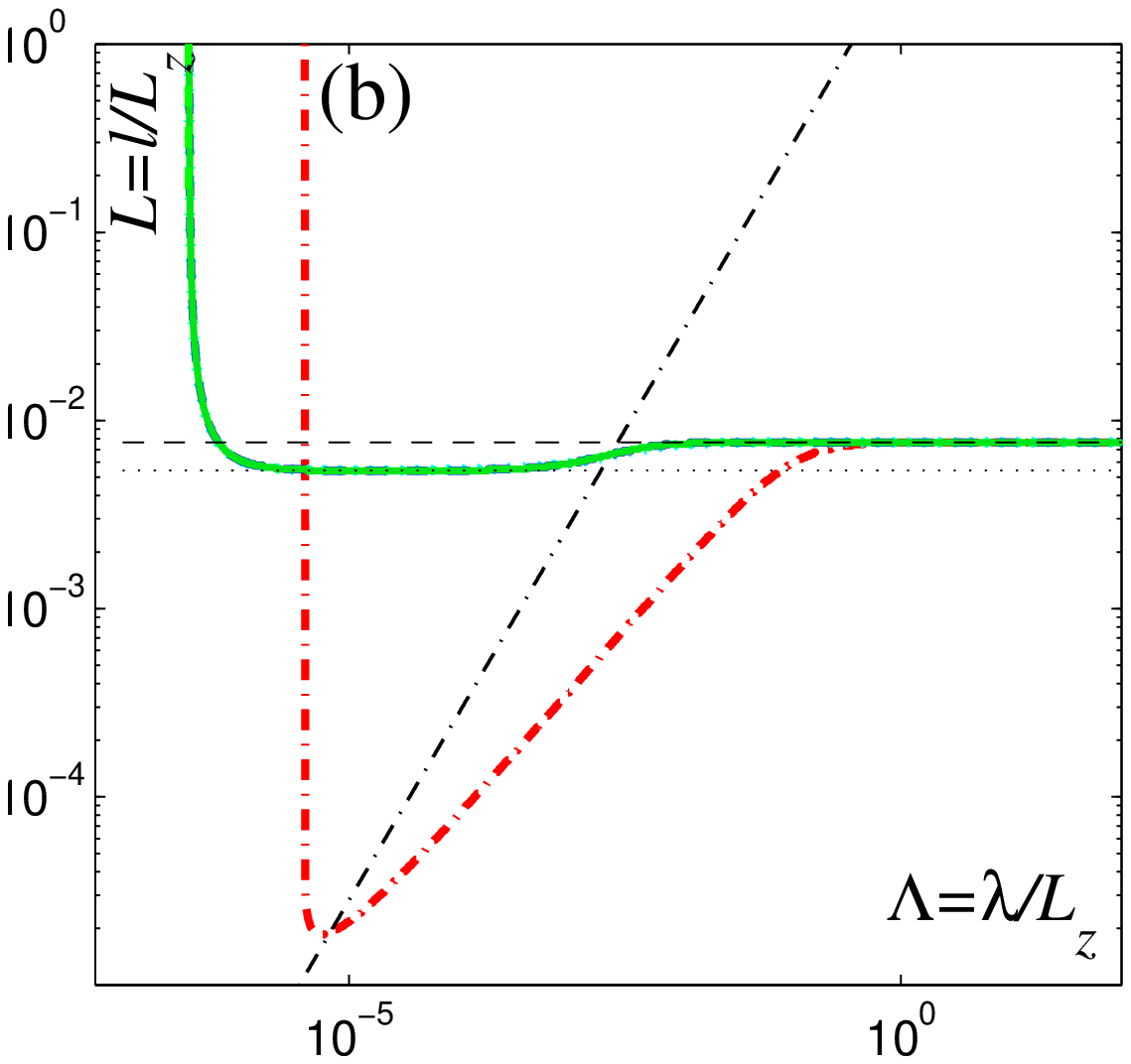}
}
\caption{Domain size $l$ normalized to the film thickness $L_z$ at equilibrium state vs the penetration depth $\Lambda=\lambda/L_z$. The solid green line corresponds to the DS which realizes the minimum of the total energy, while the dashed blue (dash-dotted red) line corresponds to the numerically calculated DS in the vortex (Meissner) state. The dotted cyan line is the analytical DS in the vortex state given by \eqref{L_S^v_L<<1}. The thin dashed, dotted and dash-dotted straight lines correspond to the normal DS value $l=l_N$, to the minimum DS value in the vortex state $l=l_N/\sqrt{2}$ and to the low temperature limit of the vortex penetration depth $l_v=\lambda/\sqrt{8}$, respectively.
The parameters are following (a) $\tilde w/L_z=10^{-4}$, $l_v/L_z=1.5\cdot 10^{-5}$ and (b) $\tilde w/L_z=10^{-4}$, $l_v/L_z=10^{-6}$.}
\label{Fig:SCferro_Lmin_and_E_SV_W<<1_Lv<Lmd_c}
\end{figure}

Keeping the main terms in \eqref{E_S*_L<<1} and \eqref{E_v*_L<<1} one can see that the penetration depth value $\lambda_m$, where the minimum values of energies $\epsilon_S$ and $\epsilon_v$ become equal, is very close to the one $\lambda_S$ ($r_\lambda\simeq 1$), where the Meissner state turns to metastable one [see Fig.~\ref{Fig:SCferro_Lmin_and_E_SV_W<<1}(a)].
Note that strictly speaking if we consider further terms in \eqref{E_S*_L<<1} and \eqref{E_v*_L<<1} we can conclude that with increasing temperature firstly the rare vortex lattice penetrates the Meissner state at $r_\lambda=1$ and after that at a bit higher temperature (corresponding to $r_\lambda\lesssim 1$, see dashed black line in Fig.~\ref{Fig:SCferro_Lmin_and_E_SV_W<<1}(a)) the global energy minimum realizes at the state with the dense vortex lattice.

 One can sum expressions \eqref{E_MH} and \eqref{E_nV+E_MH} numerically and obtain the following parameters of the phase transition: (a) the $\lambda$-dependent domain size at equilibrium which realizes the minimum of the total energy abruptly changes at $\lambda=\lambda_m$ from the vortex state one \eqref{L_S*^v_L<<1,Lmd} to the Meissner one \eqref{L_S*} [see Fig.~\ref{Fig:SCferro_Lmin_and_E_SV_W<<1}(a)]; (b) comparing the $L$~-~dependencies of the total energies in the vortex and the Meissner state for $\lambda=\lambda_m$ [see Fig.~\ref{Fig:SCferro_Lmin_and_E_SV_W<<1}(b)] one can see that $\lambda_S\simeq\lambda_m$. The essential difference of the DS in the vortex and Meissner states gives us a hint that in the case of small domain wall widths $\tilde w\ll L_z$ and for moderate magnetization values $M_c\ll M_0\ll M_{th*}$ the transition between the vortex and the Meissner states vs London penetration depth $\lambda$ (or equivalently vs magnetization amplitude $M_0$) is type I phase transition.

Note that the decrease of analytical (dotted pink) plot $l_S^v(\lambda)$ in the Fig.~\ref{Fig:SCferro_Lmin_and_E_SV_W<<1}(a) for $\lambda$ lower than  $\lambda_m$ relates to the escape of the vortices from the vortex state and to the restoration of the Meissner DS for rather rare vortex lattice.

(ii)
The scenario is very similar for the magnetization of order of critical one $M_0\lesssim M_c$, when the phase transition occurs near the minimum values of $l_{S}(\lambda)$ [see Fig.~\ref{Fig:SCferro_Lmin_and_E_SV_W<<1_Lv<Lmd_c}(a)] and therefore instead of decrease of $l_S^v(\lambda)$ with decreasing $\lambda<\lambda_m$ one can observe the increasing vortex state domain size (blue dashed line) tending to the rather large value of $l_S(\lambda)\gg L_z$ (red dash-dotted line under the solid green line).

Far from the vortex penetration threshold in this case the domain size in the vortex state
\begin{gather*}
l_S^v=l_N\sqrt{b} \ ,
\end{gather*}
is of order of the normal state one $l_N$ and it shrinks to the value $l_N/\sqrt{2}$ at the penetration depth values $\lambda$ of order of $l_N$.

(iii)
For even stronger magnetization amplitudes $M_0>M_c$ the only vortex state is stable in multidomain case ($l\ll L_x$) in the sample and the domain size $l=l_S^v$ in this case
\begin{gather*}
l_S^v=\frac{l_N\sqrt{b}}{(1-\alpha)} \ ,
\end{gather*}
behaves similar to the previous case. For $\lambda\sim l_N$ the domains also shrink to $l_N/\sqrt{2}$ and keep intact
till the divergence at the threshold $H_{c1}(\Lambda)=4\pi M_0$, due to limit $\alpha\to 1$ [see Fig.~\ref{Fig:SCferro_Lmin_and_E_SV_W<<1_Lv<Lmd_c}(b)].

As a result we can see that for the thick samples the vortex matter can essentially change the domain size comparing with the Meissner state one. The transition between these two states is proved to be type I phase transition.

\subsection{Thin films with $L_z\ll l_N$.}
Similarly to the case of the thick FS samples discussed above we present the known results for the Meissner state.
In the vicinity of the superconducting phase transition, when the effective penetration depth $\lambda^*$ is large compared with domain size $l$, the Meissner state free energy takes the form\cite{Buzd_Dao}
\begin{multline}\label{E_MH_1<<L<<Lmd}
\epsilon_S=\frac{\tilde w}{32\pi l}+\frac{1}{8}-\frac{\ln(L/\pi)+3/2}{2\pi L}+\\+
\frac{1}{\Lambda^2}\left[\frac{1}{8}-\frac{2\ln(L/\pi)}{3\pi L}\right]
\end{multline}
and reaches its minimum value
\begin{gather}\label{E_S_1<<L<<Lmd}
\epsilon_S(L_S)
\approx \frac{1}{8}\left(1+\Lambda^{-2}\right)-\frac{1}{2\pi L_N}\ ,
\end{gather}
with the equilibrium DS
\begin{multline}\label{L_S_1<<L<<Lmd}
l_S=L_z\exp\left[1+\frac{\tilde w/16 L_z-3/2+\ln\pi}{1+4L_z^2/3\lambda^2}\right]\approx \\ \approx
l_N\left(1-\frac{\tilde w L_z}{12\lambda^2}\right) \ .
\end{multline}
Note that $l_S\lesssim l_N$.

In the other limiting case of $L_z,\lambda\ll l$ the total free energy in the Meissner state is following:\cite{Buzd_Dao}
\begin{multline}\label{E_MH_1,Lmd<<L}
\epsilon_{S}\approx \frac{1}{8} + \frac{1}{4\pi l}\left(\frac{\tilde w}{8}-\pi\lambda+2 I(\lambda)\right)+\frac{\pi \lambda^4}{12 L_z l^3}\approx \\ \approx
\frac{1}{8} + \frac{1}{2\pi L}\left(\frac{\tilde w}{16 L_z}-\ln\frac{\Lambda^2}{2}-\frac{11}{12}\right)+\frac{\pi \Lambda^4}{12 L^3}\ ,
\end{multline}
where we use the expression for $I(\Lambda)$ given by \eqref{I(Lmd)} for $\lambda,\tilde w\gg L_z$. Here mentioned above critical penetration depth takes the form $\lambda_c = \sqrt{2}L_z\exp(\tilde w/32 L_z-11/24)=\sqrt{2 l_N L_z/\pi}\exp(-5/24)$, i.e. $\lambda^*_c\sim l_N$.

The minimization of free energy \eqref{E_MH_1,Lmd<<L} leads to
\begin{gather}\label{E_S_1,Lmd<<L}
\epsilon_S(L_{S})\approx \frac{1}{8}-\frac{\pi\lambda^4}{6 L_z l_S^3}
\end{gather}
reaching at the large domain width \eqref{L_S_1,Lmd<<L}
\begin{gather}\label{L_S_1<<Lmd<<L}
l_S=\frac{\pi \lambda^2}{2L_z\sqrt{\ln(\lambda/\lambda_c)}} \ .
\end{gather}
Note that $l_S\gg \lambda^*,L_z$, therefore the latter expressions work well for $\lambda/\lambda_c-1\ll 1$.
As in the previous subsection for $\lambda<\lambda_c$ the FS film goes to the monodomain state ($l\to L_x\gg L_z,\lambda^*$ in considering case) \cite{Sonin,Buzd_Faure,Buzd_Dao}.

\begin{figure}[t]
\centerline{
\includegraphics[width=0.9\linewidth]{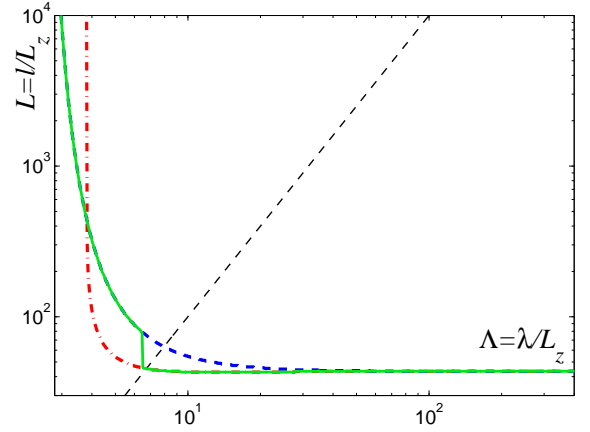}
%}
%\centerline{
%\includegraphics[width=0.88\linewidth]{Fig5b_F_SV_vs_L_eq_min_Lmd_m_Lv5p3_W50.eps}
%\includegraphics[width=0.3\linewidth]{Fig5c_F_SV_vs_L_Lmd_S_Lv5p3_W50.eps}
%\includegraphics[width=0.3\linewidth]{Fig5d_F_SV_vs_L_Lmd_V_Lv5p3_W50.eps}
}
\caption{Domain size $l$ at equilibrium state vs the penetration depth $\lambda$. The solid green line corresponds to the DS which realizes the minimum of the total energy, while the dashed blue (dash-dotted red) line corresponds to the vortex (Meissner) state.
The black dashed line corresponds to $l=\lambda^*$ and its intersection with the green line demonstrates that the phase transition occurs at $l\simeq \lambda^*$.
%
%The London penetration depths, where the state which realized total energy minimum changes ($\Lambda_m$), where the Meissner [vortex] state becomes unstable from the metastable state ($\Lambda_S$ [$\Lambda_V$]), and where the DS is diverged ($\Lambda_c$), are shown by arrows.
%
% (b-d) Comparative plots of the total energies in the vortex (solid blue line) and the Meissner (dash-dotted red line) states against the domain size $L$ (b) for $\Lambda=\Lambda_m$.%, (c) for $\Lambda=\Lambda_S$, and (d) for $\Lambda=\Lambda_V$.
%For all plots (a-d)
The parameters are following $\tilde w/L_z=50$ and $l_v/L_z=5.3$.}
\label{Fig:SCferro_Lmin_and_E_SV_W>>1}
\end{figure}

The vortices can penetrate the thin sample with the certain domain size $l$ at the critical magnetization given by \eqref{M_th_L_z,lmd_gg_l} or \eqref{M_th_lmd_ll_l} depending on the ratio $\lambda^*/l$.
Due to the fact that the shrinking of the DS in the Meissner state is very weak, see \eqref{L_S_1<<L<<Lmd}, and the crossover between the thresholds \eqref{M_th_L_z,lmd_gg_l} and \eqref{M_th_lmd_ll_l} occurs at the penetration depth $\lambda^2\sim l L_z$ of order of the critical penetration depth $\lambda_c$,  one can estimate the vortex penetration threshold into the Meissner state as $l_v^{th*}(l_N)=(4/\pi)\sqrt{2 G l_N L_z}$ \eqref{M_th_L_z,lmd_gg_l} for all multidomain states ($\lambda>\lambda_c$).

While $\lambda_c\sim(L_z l_N)^{1/2}\gg L_z$ we can restrict our consideration to the case $\lambda\gg L_z$, where the expression \eqref{E_nV+E_MH_1<<L,Lmd} for the volume energy in the vortex state works. In this case the minimum value of the total energy
\begin{gather}\label{E_v_1,Lmd<<L}
\epsilon_V(l_S^v)\approx\frac{1}{8}-\frac{(1-\tilde\alpha)^2}{2\pi L_S^v}
\end{gather}
reached at the domain width
\begin{gather}\label{L_S^v_1,Lmd<<L}
l_S^v\approx\frac{\pi L_z}{\sqrt{e}}\exp\left[\frac{\tilde w}{16 L_z(1-\tilde\alpha)^2}\right] \ .
\end{gather}
For further consideration it is useful to note that far from the vortex stability threshold $\alpha\ll 1$ the minimum energy value reduces to
\begin{gather}\label{E_v_1<<L<<Lmd_small_alpha}
\epsilon_V(l_S^v)\approx \frac{1}{8}-\frac{\exp[-\alpha \tilde w/8L_z]}{2\pi L_N}
\end{gather}
with the domain size value
\begin{gather}\label{L_S^v_1<<L<<Lmd_small_alpha}
l_S^v\approx l_N\exp\left[\frac{\alpha \tilde w}{8 L_z}\right]\ .
\end{gather}
Note that the vortex state DS is larger than the normal state value $l_S^v\gtrsim l_N$.

Let's consider firstly the vicinity of the superconducting phase transition, when $\lambda^*\gg l_N$ ($\lambda\gg\lambda_c$).
After comparison of Eqs.~(\ref{E_S_1<<L<<Lmd}, \ref{E_v_1<<L<<Lmd_small_alpha}) one can come to conclusion that the dense vortex lattice is stable only for rather strong magnetization
\begin{gather}\label{M_c}
M_0>M_c=M_{th}^*(l_N)\frac{G\tilde w}{4 \pi^3L_z} \ ,
\end{gather}
which is essentially larger than the critical value $M_{th}^*(l_N)$ by the factor $\sim\tilde w/L_z$.
Note that this critical magnetization $M_c$ is independent on $\lambda$ and consequently on the temperature.
%for any magnetization amplitudes larger than the threshold value $M_0>M_{th}^*(l_N)$ in the normal state, because the minimum value of the total energy $\epsilon_v$ in the vortex state \eqref{E_v_1<<L<<Lmd_small_alpha} formally becomes smaller than $\epsilon_S$ \eqref{E_S_1<<L<<Lmd} at even weaker magnetization

As a result for magnetization amplitudes stronger than the threshold one
\begin{gather*}
M_0>M_{th}^*(l_N) = \frac{\Phi_0\ln(l_N/\xi)}{64 K_t L_z l_N}\sim \frac{\Phi_0 \tilde w}{L_z^3}\exp\left[-\frac{\tilde w}{32 L_z}\right]  \ ,
\end{gather*}
the global minimum of the total energy seems to be realized in the vortex state with rare lattice, while the Meissner state doesn't realize even local minimum of energy and the state with the dense vortex lattice have the larger energy in this case.
Note that the Meissner state is unstable to formation of the rare vortex lattice even for $\lambda<\lambda_c$ due to $M_{th}^\lambda(\lambda_c)\simeq M_{th*}(l_N)$. Thus, as was mentioned in the Sec.~\ref{sec4_eq_vortex_density} our model fails in the case of rather thin FS films with large values of the penetration depth $\lambda^*\gg l_N$, due to rude approximation for the vortex density $n_{eq}(x)$.
%only the vortex state is stable for any $\lambda\gg\lambda_c$, due to $l_S\gtrsim l_N$.

Further we will assume that rather rare vortex lattice doesn't change crucially the magnetic domain structure and will use the Meissner energy values as an approximation of the energy of rare vortex lattice
to estimate the parameter values at the transition between rare and dense vortex states.

%Let's consider the case of the magnetization values weaker than the critical magnetization $M_0\lesssim M_c$, when for $\lambda> \lambda_c$ the phase transition between Meissner and the vortex states may occur.
%%Let's consider the case of the magnetization values weaker than the critical magnetization in the normal state $M_0\lesssim M_{th}^*(l_N)$, when for $\lambda\gg \lambda_c$ the phase transition between Meissner and the vortex states may occur.
%%As it was mentioned above the dense lattice (if any vortex can penetrate the sample) is more energetically favorable than the Meissner one for rather low magnetization values \eqref{M_c}.
%%Therefore to the purpose of realization of the vortex state we need that the vortex DS for general $\alpha$ values \eqref{L_S^v_1,Lmd<<L} will be essentially larger than the normal state value $l_N$.
%%More carefully we can write that
%%$$l_v^2<\frac{32 G l_S^v L_z}{\pi^2}=\frac{32 G L_z^2}{\pi}\exp\left[\frac{\tilde w}{16(1-l_v^2/8\lambda^2)^2}\right] \ .$$
%%Evaluating the latter inequality one can see that such a transition from Meissner to vortex state with decreasing temperature may occur only in the vicinity of the vortex stability threshold (see Fig.~\ref{Fig:SCferro_Lmin_and_E_SV_W>>1}):
%%$$\left(1-\frac{l_v^2}{8\lambda^2}\right)^2\approx \frac{\tilde w}{16 L_z \ln(\lambda^2/4\pi G L_z^2)}\ll 1 \ .$$
%In general case for $\lambda^2\gtrsim l_N L_z$ the dense vortex lattice becomes more energetically favorable at $\lambda>\lambda_c$, when

%
% [see Fig.~\ref{Fig:SCferro_Lmin_and_E_SV_W>>1_Lv<Lmd_c}(a)].

In the vicinity of the critical $\lambda_c$ one can
compare \eqref{E_S_1,Lmd<<L} with $\lambda\simeq\lambda_c$
$$\epsilon_S(l_S)\approx \frac{1}{8}-\frac{2 e^{5/12}}{3 \pi L_N} \left[\ln(\lambda/\lambda_c)\right]^{3/2}$$
 and \eqref{E_v_1,Lmd<<L} and conclude that the dense vortex lattice can realize the global minimum only for $M_0\gtrsim M_c\ln[\ln(\lambda/\lambda_c)^{-3/2}]\gg M_{th}^*(L_N)$, which is even larger than the critical magnetization at $\lambda^*\gg l_N$.

As a result we prove for thin FS samples $L_z\ll \tilde w$ that in the ranges $\lambda\simeq \lambda_c$ and $\lambda^2\gg l_N L_z$ the state with the dense vortex lattice can realize the global minimum of the total energy for rather strong magnetization values $M_0>M_c$ and this state remains to have the global minimum energy value till the monodomain state.
Unfortunately for weaker magnetization values our model fails to describe the inhomogeneous vortex lattice which is rather rare in the middle of the domains.

Finally, for $M_0\lesssim M_c$ we consider numerically the middle region of temperatures, when $\lambda^2\sim l_N L_z$ and $\lambda/\lambda_c-1\gtrsim 1$, where our analytical approximations fail. One can see from Fig.~\ref{Fig:SCferro_Lmin_and_E_SV_W>>1} that in this case type I phase transition from the Meissner (read ``rare vortex lattice state'') to the vortex state with dense lattice occurs in this region.

\section{Discussion}\label{sec6_Conclusion}
 Within the model described above we have investigated the equilibrium vortex density distribution and the equilibrium state of the ferromagnetic superconductor sample with stripe-structured magnetic domains and have found that the type I phase transitions between the Meissner and the vortex states occur in the sample with decreasing temperature.

For describing the vortex state we use a continuous model, which is valid for the dense vortex lattices and not so close to the vortex penetration threshold.
In the above-mentioned restrictions our consideration gives the best results for rather thick samples $L_z\gg\tilde w$ in the range of moderate penetration depths $\tilde w\ll\lambda\ll\tilde w^{1/4}L_z^{3/4}$, when domains shrink strongly in the Meissner state (see Fig.~\ref{Fig:SCferro_Lmin_and_E_SV_W<<1}).\cite{Buzd_Faure} In this case the validness of our consideration near the vortex threshold in the Meissner state is provided by a very large ratio $l_{S*}/l_S^v \gg 1$ of the domain size in the Meissner $l_{S*}$ and the vortex $l_{S}^v$ states.

Throughout the paper
%In addition
we neglected all the pinning potentials both for the domain walls and for the vortices.
However, we can easily take into account strong domain pinning potential (neglecting the vortex pinning effects) by considering the energy for fixed domain size $l$ without minimization over $l$.
For many experimental situations it is the typical case that the magnetic domains pinning strongly overcome the vortex pinning. In such a case in each ferromagnetic domain the vortex concentration should be equal to an equilibrium one.
It is interesting to note that applying rather small magnetic field $H$ along the magnetization ($H\parallel Oz$) one can obtain paramagnetic response of the vortices.
Indeed, for the domain with codirectional the magnetization and the external field the total internal field is $4\pi M+H$ and the diamagnetic contribution to the moment equals to $4\pi\delta M^{c}=-\Phi_0/(8\pi\lambda^2)[\ln(\eta H_{c2}^{c}/(4\pi M+H))]$, (see, e.g., [\onlinecite{AladMoshch-review,deGennes}]), where $\eta$ is the demagnetization coefficient and $H_{c2}$ is the upper critical field along $z$ axis. For the domains with an opposite orientation of magnetization the internal field is $H-4\pi M$, i.e. it is opposite to the weak applied field and the corresponding diamagnetic moment $4\pi\delta M^{c}=\Phi_0/(8\pi\lambda^2)[\ln(\eta H_{c2}^{c}/(4\pi M-H))]$. Assuming that the applied field $H$ is smaller than the coercive field the size of the up and down domains should be the same and the averaged magnetic moment due to the vortices is positive $4\pi\left<\delta M^{c}\right>=+\Phi_0/(8\pi \lambda^2)H/(4\pi M)$, for $H\ll 4\pi M$. Note that for the fields higher than the coercive field the size of the ``parallel'' domains should grow and this would decrease the paramagnetic response and eventually switch it to the diamagnetic one.

On experiments [\onlinecite{UCoGe-2_Paulsen-2012}] in UCoGe the diamagnetic response was observed and this may be explained by the strong vortex pinning - the vortex configuration is frozen and only the surface Meissner current contributes to the screening.

\acknowledgements
We thank Prof.~J.~Flouquet for useful discussions.
This work was supported, in part,
by European IRSES program SIMTECH (contract n.246937), French ANR project ``ElectoVortex'', NanoSC COST Action MP1201, the Russian Foundation
for Basic Research, FTP Scientific and educational personnel of innovative Russia in
2009-2013, and the Russian president foundation (SP-1491.2012.5).

\end{document}